\documentclass[twoside,12pt]{article}
\usepackage{epsfig}

\newcommand{\be}{\begin{equation}}
\newcommand{\ee}{\end{equation}}
\newcommand{\bea}{\begin{eqnarray}}
\newcommand{\eea}{\end{eqnarray}}

\usepackage{amsmath}
\usepackage{amsfonts}
\usepackage{amssymb}
\usepackage{amsxtra}
\usepackage{supertabular}
\begin{document}
\title{Clifford odd and even objects offer description of internal space of fermions
and bosons, respectively, opening new insight into the second quantization of fields}

\author{N.S.\ Manko\v c Bor\v stnik,$^{1}$\\
$^1$Department of Physics, University of Ljubljana\\
SI-1000 Ljubljana, Slovenia\\
norma.mankoc@fmf.uni-lj.si}
\maketitle



\begin{abstract}
In a long series of works the author has demonstrated that the model named the 
{\it spin-charge-family} theory offers the explanation for all in the {\it standard model}
assumed properties of the fermion and boson fields, as well as for many of their so far 
observed properties if the space-time is $\ge (13 +1)$ while fermions interact
with gravity only. In this talk, I briefly report on the so far achievements of the theory. 
The main contribution demonstrates the offer of the Clifford odd and even objects  
for the description of the internal spaces of fermion (Clifford odd) and boson (Clifford 
even) fields, which is opening up a new understanding of the second quantization 
postulates for the fermion and boson fields: The "basis vectors" determined by the 
Clifford odd objects demonstrate all the properties of the internal space of fermions 
and transfer their anticommutativity to their creation and annihilation operators, 
while the "basis vectors" determined by the Clifford even objects demonstrate all 
the properties of the internal space of boson fields and transfer their commutativity 
to their creation and annihilation operators. The toy model with $d=(5+1)$ illustrates
the statements.
\end{abstract}

\section {Introduction}

\vspace{2mm}

The {\it standard model} (improved with neutrinos having non zero masses) has 
been experimentally confirmed without raising any serious doubts so far on its 
assumptions, which remain unexplained, waiting for the explanation. 

The {\it standard model}  has in the literature several  explanations, mostly 
with many new not explained assumptions. The most  popular seem to be 
the grand unifying theories~(\cite{Geor,FritzMin,PatiSal,GeorGlas,Cho,ChoFreu} and 
many others.

Among the questions which must be answered are:\\
{\bf i.} Where do fermions, quarks and leptons,  with their families included originate 
and why do their internal spaces differ so much from the internal spaces of boson fields? \\
{\bf ii.} Why are charges of quarks and leptons so different and why have the 
left handed family members so different charges from the right handed ones? \\
{\bf iii.} Where do antiquarks and antileptons originate?\\
{\bf iv.} Where do families of quarks and leptons originate and how many 
families do exist? \\
{\bf v.} Why do family members -- quarks and leptons --- manifest so different 
masses if they all start as massless?\\
{\bf vi.} What is the origin of boson fields,  of vector fields which are the gauge fields 
of fermions, and of the Higgs's scalars and the Yukawa couplings.\\
{\bf vii.} How are scalar fields connected with the origin of families and how many 
scalar fields determine properties of the so far (and others possibly be) observed 
fermions and masses of weak bosons? \\
{\bf viii.} Do possibly exist also scalar fields with the  colour charges in the 
fundamental representation (like the Higgs's scalars are doublets with respect 
to the weak charge) and where, if they are, do they manifest?\\
{\bf  ix.}  Can it be that all boson fields, with the scalar fields, included, have a 
common  origin? \\
{\bf x.} Where does the {\it dark matter} originate?\\
{\bf xi.}  Where does the "ordinary"  matter-antimatter asymmetry originate?\\
{\bf xii.} Where does the  dark energy originate and why is it so small?\\
{\bf xiii.} How can we understand the second quantized fermion and boson fields?\\
{\bf xiv.} What is the dimension of space? $(3+1)?$, $((d-1)+1)?$, $\infty?$ \\
{\bf xv.} Are all the fields indeed second quantized with the gravity included?
And consequently all the systems are second quantized (although we can treat them
in simplified versions, like it is the first quantization and even the classical treatment),
with the black holes included?
{\bf xvi.} And many others.

In a long series of works~(\cite{norma92,norma93,norma95,IARD2016,n2014matterantimatter,%
nd2017,n2012scalars,JMP2013,normaJMP2015,nh2018} and the references therein), the 
author has succeeded,  together with collaborators, to find the answer to the above, and
also other open questions of the {\it standard model} and also to several open 
cosmological questions, with the model named the {\it spin-charge-family} theory. 

The theory assumes that the space has more than $(3+1)$ dimensions, it must 
have $d\ge (13+1)$, so that the subgroups of the $SO(13,1)$ group, describing 
the internal space of fermions by the superposition of odd products of the Clifford 
objects $\gamma^{a}$'s, manifest from the point of view of 
$d=(3+1)$-dimensional space the spins, handedness and charges assumed for 
massless fermions in the {\it standard model}. 
Correspondingly each irreducible representation of the $SO(13,1)$ group carrying 
the quantum numbers of quarks and leptons and antiquarks and antileptons, 
represents one of families of fermions, the quantum numbers of which are 
determined by the second kind of the Clifford objects, by  $\tilde{\gamma}^a $
(by $\tilde{S}^{ab}$ $(=\frac{i}{4}\,\{\tilde{\gamma}^a,\, 
\tilde{\gamma}^{b}\}_{-}$, indeed).

Fermions interact in $d=(13+1)$ with gravity only, with vielbeins (the gauge fields of 
momenta) and the two kinds of the spin connection fields, the gauge fields of the two 
kinds of the Lorentz transformations in the internal space of fermions, of $S^{ab}=
\frac{i}{4}\,\{\gamma^a,\, \gamma^{b}\}_{-}$ and of $\tilde{S}^{ab}$ 
$(=\frac{i}{4}\,\{\tilde{\gamma}^a,\, \tilde{\gamma}^{b}\}_{-}$.

The theory assumes a simple starting action~(\cite{nh2021RPPNP} and the references 
therein)  for the second quantized massless fermion and antifermion fields, and the 
corresponding massless boson fields in  $d=2(2n+1)$-dimensional space 
\begin{eqnarray}
{\cal A}\,  &=& \int \; d^dx \; E\;\frac{1}{2}\, (\bar{\psi} \, \gamma^a p_{0a} \psi) 
+ h.c. +
\nonumber\\  
               & & \int \; d^dx \; E\; (\alpha \,R + \tilde{\alpha} \, \tilde{R})\,,
\nonumber\\
               p_{0a } &=& f^{\alpha}{}_a p_{0\alpha} + \frac{1}{2E}\, \{ p_{\alpha},
E f^{\alpha}{}_a\}_- \,,\nonumber\\
          p_{0\alpha} &=&  p_{\alpha}  - \frac{1}{2}  S^{ab} \omega_{ab \alpha} - 
                    \frac{1}{2}  \tilde{S}^{ab}   \tilde{\omega}_{ab \alpha} \,,
                    \nonumber\\                    
R &=&  \frac{1}{2} \, \{ f^{\alpha [ a} f^{\beta b ]} \;(\omega_{a b \alpha, \beta} 
- \omega_{c a \alpha}\,\omega^{c}{}_{b \beta}) \} + h.c. \,, \nonumber \\
\tilde{R}  &=&  \frac{1}{2} \, \{ f^{\alpha [ a} f^{\beta b ]} 
\;(\tilde{\omega}_{a b \alpha,\beta} - \tilde{\omega}_{c a \alpha} \,
\tilde{\omega}^{c}{}_{b \beta})\} + h.c.\,.               
\label{wholeaction}
\end{eqnarray}
Here~\footnote{$f^{\alpha}{}_{a}$ are inverted vielbeins to 
$e^{a}{}_{\alpha}$ with the properties $e^a{}_{\alpha} f^{\alpha}{\!}_b = 
\delta^a{\!}_b,\; e^a{\!}_{\alpha} f^{\beta}{\!}_a = \delta^{\beta}_{\alpha} $, 
$ E = \det(e^a{\!}_{\alpha}) $.
Latin indices  
$a,b,..,m,n,..,s,t,..$ denote a tangent space (a flat index),
while Greek indices $\alpha, \beta,..,\mu, \nu,.. \sigma,\tau, ..$ denote an Einstein 
index (a curved index). Letters  from the beginning of both the alphabets
indicate a general index ($a,b,c,..$   and $\alpha, \beta, \gamma,.. $ ), 
from the middle of both the alphabets   
the observed dimensions $0,1,2,3$ ($m,n,..$ and $\mu,\nu,..$), indexes from 
the bottom of the alphabets
indicate the compactified dimensions ($s,t,..$ and $\sigma,\tau,..$). 
We assume the signature $\eta^{ab} =
diag\{1,-1,-1,\cdots,-1\}$.} 
$f^{\alpha [a} f^{\beta b]}= f^{\alpha a} f^{\beta b} - f^{\alpha b} f^{\beta a}$.
$f^a_{\alpha}$, and the two kinds of the spin connection fields, 
$\omega_{ab \alpha}$ (the gauge fields of $S^{ab}$) and $\tilde{\omega}_{ab \alpha}$  
(the gauge fields of $\tilde{S}^{ab}$), manifest in $d=(3+1)$ as the known vector 
gauge fields and the scalar gauge fields taking care of masses of quarks and leptons and antiquarks and antileptons and the weak boson fields~\cite{nd2017}~\footnote{
Since the multiplication with either $\gamma^a$'s or $\tilde{\gamma}^a$'s  changes 
the Clifford odd ''basis vectors'' into the Clifford even  objects, 
and even ''basis vectors'' commute, the action for fermions can not include an odd 
numbers of $\gamma^a$'s or $\tilde{\gamma}^a$'s, what the simple starting action 
of Eq.~(\ref{wholeaction}) does not. In the starting action $\gamma^a$'s and 
$\tilde{\gamma}^a$'s appear as $\gamma^0 \gamma^a \hat{p}_a$  or as 
$\gamma^0 \gamma^c \, S^{ab}\omega_{abc}$  and  as 
$\gamma^0 \gamma^c \,\tilde{S}^{ab}\tilde{\omega}_{abc} $.}

While in any even dimensional space the superposition of odd products of 
$\gamma^{a}$'s, forming the Clifford odd ''basis vectors'', offer the description of 
the internal space of fermions with the half integer spins, 
the superposition of even products of $\gamma^a$'s, forming the Clifford even ''basis 
vectors'', offer the description  of the internal space of boson fields with integer spins, 
manifesting as gauge fields of the corresponding Clifford odd ''basis vectors''.

From the point of view of $d=(3+1)$ one family of the Clifford odd ''basis vectors'' 
with $2^{\frac{d}{2}-1}$ members manifest spins, handedness and charges of quarks 
and leptons and antiquarks and antileptons appearing in $2^{\frac{d}{2}-1}$ families, 
while their Hermitian conjugated partners appear in another group of $2^{\frac{d}{2}-1}$
members in $2^{\frac{d}{2}-1}$ families. 

The Clifford even ''basis vectors'' appear in two groups, each with $2^{\frac{d}{2}-1}
\times 2^{\frac{d}{2}-1}$ members, with the Hermitian conjugated partners within
the same group and have correspondingly no families. The Clifford even ''basis vectors''
manifest from the point of view of $d=(3+1)$ all the properties of the vector gauge 
fields before the electroweak break and for the scalar fields  causing the electroweak 
break as assumed by the {\it standard model}. 

Tensor products of the Clifford odd and Clifford even  ''basis vectors'' with the basis in 
ordinary space form the creation operators to which the ''basis vectors'' transfer either
anticommutativity  or commutativity. The Clifford odd ''basis vectors'' transfer their 
anticommutativity to creation operators and to their Hermitian conjugated partners 
annihilation operators. The Clifford even ''basis vectors'' transfer their commutativity
to creation operators and annihilation operators. Correspondingly the 
anticommutation properties of creation and annihilation operators of fermions explain 
the second quantization postulates of Dirac for  fermion fields, while the commutation 
properties of creation and annihilation operators for bosons explain the corresponding 
second quantization postulates for  boson fields.

The  {\it spin-charge-family} theory is offering answers to most of the open questions
presented above. The more work is put into the theory the more answers the theory 
offers.
 
In Sect.~\ref{creationannihilation} the Grassmann and the Clifford algebra is explained
and creation and annihilation operators described as a tensor products of the 
''basis vectors'' offering explanation of the internal spaces of fermion (by the Clifford 
odd algebra) and boson (by the Clifford even algebra) fields and the basis in ordinary
 space.

In Subsect.~\ref{basisvectors} the ''basis vectors'' are introduced and their properties 
presented.

In Subsect.~\ref{cliffordoddevenbasis5+1} the properties of the Clifford odd and even
''basis vectors'' are demonstrated in the toy model in $d=(5+1)$.

In Subsect.~\ref{secondquantizedfermionsbosons} the properties of the creation and 
annihilation operators for the second quantized fields are described.

In Sect.~\ref{SCFT} a short overview of the achievements and  predictions so far of 
the {\it spin-charge-family} theory is presented,

Sect.~\ref{conclusions} presents what the reader could learn from the main 
contribution of this talk.

\section{Creation and annihilation operators for fermions and bosons}
\label{creationannihilation}
\vspace{2mm}

The second quantization postulates for fermions~\cite{Dirac,BetheJackiw,Weinberg} 
require that the creation operators and their Hermitian conjugated partners annihilation 
operators, depending on  a finite dimensional basis in internal space, that is on the space
of half integer spins and on charges described by the fundamental representations of the 
appropriate groups, and on continuously infinite number of momenta (or coordinates)~(\cite{nh2021RPPNP}, Subsect.~3.3.1), fulfil anticommutation relations.  

The second quantization postulates for bosons~\cite{Dirac,BetheJackiw,Weinberg} 
require that the creation and annihilation operators, depending on finite dimensional 
basis in internal space, that is on the space of integer spins and on charges described 
by the adjoint representations of the same groups, and on continuously infinite number
of momenta (or coordinates)~(\cite{nh2021RPPNP}, Subsect. ~3.3.1), fulfil 
commutation relation. 

I demonstrate in this talk that using the Clifford algebra to describe the internal space
of fermions and bosons, the creation and annihilation operators which are tensor
products of the internal basis and the momentum/coordinate basis, not only fulfil the 
appropriate anticommutation  relations (for fermions) or commutation relations 
(for bosons) but also have the required properties for either fermion fields 
(if the internal space is described 
with the Clifford odd products of $\gamma^{a}$'s) or of boson fields (if the internal 
space is described with the Clifford even products of $\gamma^{a}$'s). The  Clifford 
odd and Clifford even ''basic vectors'' correspondingly offer the explanation for the 
second quantization postulates for fermions and bosons, respectively. 

There are two Clifford subalgebras which can be used to describe the internal 
space of fermions and of bosons, each with $2^d$ members. In each of the two 
subalgebras  there are 2 $\times$ $2^{\frac{d}{2}-1}\times$ $ 2^{\frac{d}{2}-1}$ 
Clifford odd and 2 $\times$ $2^{\frac{d}{2}-1}\times$ $ 2^{\frac{d}{2}-1}$ 
Clifford even ''basic vectors'' which can be used to describe the internal space of 
fermion fields, the Clifford odd ''basic vectors'', and of boson fields, the Clifford even 
''basic vectors'' in any even $d$. 
$d=(13+1)$ offers the explanation for all the properties of  fermion fields, with 
families included, and boson fields which are the gauge fields of fermion fields. 

In any even $d$, $d=2(2n+1)$ or $d=4n$, any of the two Clifford subalgebras 
offers  twice $2^{\frac{d}{2}-1}$ irreducible representations, each with 
$2^{\frac{d}{2}-1}$ members, which can represent ''basis vectors''and their 
Hermitian conjugated partners. Each irreducible representation offers in 
$d=(13+1)$ the description of the quarks and the antiquarks and the leptons 
and the antileptons (with the right handed neutrinos and left handed antineutrinos
included in addition what is) assumed by the {\it standard model}.

There are obviously only one kind of fermion fields and correspondingly also of 
their gauge fields observed. There is correspondingly no need for two Clifford 
subalgebras.

The reduction of the two subalgebras to only one with the postulate in Eq.~~(\ref{tildegammareduced}), (Ref.~\cite{nh2021RPPNP}, Eq.~(38)) solves 
this problem. At the same time the reduction offers the quantum numbers for each
of the irreducible representations of the Clifford subalgbebra left, $\gamma^{a}$'s, 
when fermions are concerned~(\cite{nh2021RPPNP} Subsect.~3.2).

Boson fields have no families as it will be demonstrated.\\

\vspace{2mm}

$\;\;\;$ {\it Grassmann and Clifford algebras}

\vspace{2mm}

The internal space of anticommuting or commuting second quantized fields can be 
described by using either the Grassmann or the Clifford algebras~\cite{norma92,%
norma93,norma95,normaJMP2015}.  What follows is a short overview of Subsect.3.2
of Ref.~\cite{nh2021RPPNP} and of references cited in~\cite{nh2021RPPNP}.
 
In Grassmann $d$-dimensional space there are $d$ anticommuting (operators) 
$\theta^{a}$,
 and 
$d$ anticommuting operators which are derivatives with respect to $\theta^{a}$,
$\frac{\partial}{\partial \theta_{a}}$, 
%
\begin{eqnarray}
\label{thetaderanti0}
\{\theta^{a}, \theta^{b}\}_{+}=0\,, \, && \,
\{\frac{\partial}{\partial \theta_{a}}, \frac{\partial}{\partial \theta_{b}}\}_{+} =0\,,
\nonumber\\
\{\theta_{a},\frac{\partial}{\partial \theta_{b}}\}_{+} &=&\delta_{ab}\,, (a,b)=(0,1,2,3,5,\cdots,d)\,.
\end{eqnarray}

Defining~\cite{nh2018} 
\begin{eqnarray}
(\theta^{a})^{\dagger} &=& \eta^{a a} \frac{\partial}{\partial \theta_{a}}\,,\quad
{\rm leads  \, to} \quad
(\frac{\partial}{\partial \theta_{a}})^{\dagger}= \eta^{a a} \theta^{a}\,,
\label{thetaderher0}
\end{eqnarray}
with $\eta^{a b}=diag\{1,-1,-1,\cdots,-1\}$.

$ \theta^{a}$ and $ \frac{\partial}{\partial \theta_{a}}$ are, up to the sign, Hermitian conjugated to each other. The identity is the self adjoint member of the algebra.
The choice for the following complex properties of $\theta^a$ and correspondingly 
of $\frac{\partial}{\partial \theta_{a}}$ are made
\begin{eqnarray}
\label{complextheta}
\{\theta^a\}^* &=&  (\theta^0, \theta^1, - \theta^2, \theta^3, - \theta^5,
\theta^6,...,- \theta^{d-1}, \theta^d)\,, \nonumber\\
\{\frac{\partial}{\partial \theta_{a}}\}^* &=& (\frac{\partial}{\partial \theta_{0}},
\frac{\partial}{\partial \theta_{1}}, - \frac{\partial}{\partial \theta_{2}},
\frac{\partial}{\partial \theta_{3}}, - \frac{\partial}{\partial \theta_{5}}, 
\frac{\partial}{\partial \theta_{6}},..., - \frac{\partial}{\partial \theta_{d-1}}, 
\frac{\partial}{\partial \theta_{d}})\,. 
\end{eqnarray}

The are $2^d$ superposition of products of  $\theta^{a}$, 
the Hermitian conjugated partners of which are the corresponding superposition of products 
of $\frac{\partial}{\partial \theta_{a}}$.

There exist two kinds of the Clifford algebra elements (operators), $\gamma^{a}$ and 
$\tilde{\gamma}^{a}$, expressible with $\theta^{a}$'s and their conjugate momenta $p^{\theta a}= i \,\frac{\partial}{\partial \theta_{a}}$ ~\cite{norma93}, Eqs.~(\ref{thetaderanti0}, \ref{thetaderher0}), 
\begin{eqnarray}
\label{clifftheta1}
\gamma^{a} &=& (\theta^{a} + \frac{\partial}{\partial \theta_{a}})\,, \quad 
\tilde{\gamma}^{a} =i \,(\theta^{a} - \frac{\partial}{\partial \theta_{a}})\,,\nonumber\\
\theta^{a} &=&\frac{1}{2} \,(\gamma^{a} - i \tilde{\gamma}^{a})\,, \quad 
\frac{\partial}{\partial \theta_{a}}= \frac{1}{2} \,(\gamma^{a} + i \tilde{\gamma}^{a})\,,
\nonumber\\
\end{eqnarray}
offering together  $2\cdot 2^d$  operators: $2^d$ are superposition of products of 
$\gamma^{a}$  and  $2^d$  of $\tilde{\gamma}^{a}$.
It is easy to prove, if taking into account Eqs.~(\ref{thetaderher0}, \ref{clifftheta1}),
 that they form two anticommuting Clifford subalgebras, 
$\{\gamma^{a}, \tilde{\gamma}^{b}\}_{+} =0$, Refs.~(\cite{nh2021RPPNP} and 
references therein)
\begin{eqnarray}
\label{gammatildeantiher}
\{\gamma^{a}, \gamma^{b}\}_{+}&=&2 \eta^{a b}= \{\tilde{\gamma}^{a}, 
\tilde{\gamma}^{b}\}_{+}\,, \nonumber\\
\{\gamma^{a}, \tilde{\gamma}^{b}\}_{+}&=&0\,,\quad
 (a,b)=(0,1,2,3,5,\cdots,d)\,, \nonumber\\
(\gamma^{a})^{\dagger} &=& \eta^{aa}\, \gamma^{a}\, , \quad 
(\tilde{\gamma}^{a})^{\dagger} =  \eta^{a a}\, \tilde{\gamma}^{a}\,.
\end{eqnarray}

 \vspace{3mm}

While the Grassmann algebra offers the description of the ''anticommuting integer spin
second quantized fields'' and of the ''commuting integer spin second quantized 
fields''~\cite{n2021MDPIsymmetry,nh2021RPPNP}, the Clifford algebras which are 
superposition of odd products of either $\gamma^a$'s or $\tilde{\gamma}^a$'s offer 
the description of the second quantized half integer spin fermion fields, which from 
the point of the subgroups of the $SO(d-1,1)$ group manifest spins and charges of 
fermions and antifermions in the fundamental representations of the group and 
subgroups.

\noindent
The superposition of even products of either $\gamma^a$'s or $\tilde{\gamma}^a$'s 
offer the description of the commuting second quantized boson fields with integer spins 
(as we can see in~\cite{n2021SQ} and shall see in this contribution) which from the 
point of the subgroups of the $SO(d-1,1)$ group manifest spins and charges in the 
adjoint representations of the group and subgroups.

The following {\it postulate}, which determines how does  $\tilde{\gamma}^{a}$'s 
operate on $\gamma^a$'s,  reduces the two Clifford subalgebras, $\gamma^a$'s and $\tilde{\gamma}^a$'s, to one, to the one described by 
$\gamma^a$'s~\cite{nh03,norma93,JMP2013,normaJMP2015,nh2018}
\begin{eqnarray}
\{\tilde{\gamma}^a B &=&(-)^B\, i \, B \gamma^a\}\, |\psi_{oc}>\,,
\label{tildegammareduced}
\end{eqnarray}
with $(-)^B = -1$, if $B$ is (a function of) an odd products of $\gamma^a$'s,  otherwise 
$(-)^B = 1$~\cite{nh03}, $|\psi_{oc}>$ is defined in Eq.~(\ref{vaccliffodd}). of 
Subsect.~\ref{basisvectors}.

\vspace{2mm}

After the postulate of Eq.~(\ref{tildegammareduced}) it follows:\\
{\bf a.} The Clifford subalgebra described by $\tilde{\gamma}^{a}$'s looses its meaning 
for the description of the internal space of quantum fields.\\
{\bf b.} The ''basis vectors'' which are  superposition of an odd or an even products of 
$\gamma^a$'s obey the postulates for the second quantization fields for fermions or 
bosons, respectively, Sect.\ref{basisvectors}.\\
{\bf c.} It can be proven that the relations presented in Eq.~(\ref{gammatildeantiher})
remain valid also after the postulate of Eq.~(\ref{tildegammareduced}). The proof is
presented in Ref.~(\cite{nh2021RPPNP}, App.~I, Statement~3a.\\
{\bf d.} Each irreducible representation of the Clifford odd ''basis vectors'' described by
$\gamma^{a}$'s are equipped by the quantum numbers of  the Cartan subalgebra 
members of $\tilde{S}^{ab}$, chosen in Eq.~(\ref{cartangrasscliff}), 
 as follows
\begin{eqnarray}
&&{\cal {\bf S}}^{03}, {\cal {\bf S}}^{12}, {\cal {\bf S}}^{56}, \cdots, 
{\cal {\bf S}}^{d-1 \;d}\,, \nonumber\\
&&S^{03}, S^{12}, S^{56}, \cdots, S^{d-1 \;d}\,, \nonumber\\
&&\tilde{S}^{03}, \tilde{S}^{12}, \tilde{S}^{56}, \cdots,  \tilde{S}^{d-1\; d}\,, 
\nonumber\\
&&{\cal {\bf S}}^{ab} = S^{ab} +\tilde{S}^{ab}=
 i \, (\theta^{a} \frac{\partial}{\partial \theta_{b}} - 
 \theta^{b} \frac{\partial}{\partial \theta_{a}})\,.
\label{cartangrasscliff}
\end{eqnarray}

\vspace{2mm}

After the  postulate of Eq.~(\ref{tildegammareduced}) no vector space of 
$\tilde{\gamma}^{a}$'s needs to be taken into account for the description of 
the internal space of either fermions or bosons, in agreement with the observed 
properties of fermions and bosons. Also the Grassmann algebra is reduced to only
one of the Clifford subalgebras. The operators $\tilde{\gamma}^{a}$'s describe 
from now on properties of fermion and boson ''basis vectors'' determined by
superposition of products of odd or even numbers of $\gamma^a$'s, respectively.

$\tilde{\gamma}^a$'s  equip each irreducible representation of the Lorentz group 
(with the infinitesimal generators $S^{ab}=\frac{i}{4} \{\gamma^a, \,
\gamma^b\}_{-}$) when applying on  the Clifford odd ''basis vectors'' (which are 
superposition of odd products of $\gamma^{a's}$) with the family quantum 
numbers (determined by $\tilde{S}^{ab}=\frac{i}{4} \{\tilde{\gamma}^a, 
\,\tilde{\gamma}^b\}_{-}$). 

Correspondingly the Clifford odd ''basis vectors'' (they are superposition of an odd 
products of $\gamma^{a}$'s) form $2^{\frac{d}{2}-1}$ families, with the 
quantum number $f$, each family  have $2^{\frac{d}{2}-1}$ members, $m$. 
They offer the description of the second quantized fermion fields.

The Clifford even ''basis vectors'' (they are superposition of an even products of 
$\gamma^a$'s) have no families as we shall see in what follows, but they do 
carry both quantum numbers, $f$ and $m$.
They offer the description of the second quantized boson fields as the gauge 
fields of the second quantized fermion fields. The generators of the Lorentz 
transformations in the internal space of the Clifford even ''basis vectors'' are  
${\bf {\cal S}}^{ab}= S^{ab} + \tilde{S}^{ab}$. 
 
Properties of the Clifford odd and the Clifford even ''basis vectors'' are discussed 
in the next subsection.

\vspace{1mm}
\subsection{''Basis vectors'' of fermions and bosons}
\label{basisvectors} 

\vspace{2mm}

After the reduction of the two Clifford subalgebras to only one, 
Eq.~(\ref{tildegammareduced}), we only need to define  ''basis vectors'' for the case 
that the internal space of second quantized fields is described by superposition of odd or even products $\gamma^{a}$'s~\footnote{
In Ref.~\cite{nh2021RPPNP} the reader can find in 
Subsects.~(3.2.1 and 3.2.2) definitions for the ''basis vectors'' for the Grassmann and 
the two Clifford subalgebras, which are products of nilpotents and projectors chosen 
to be eigenvactors of the corresponding Cartan subalgebra members of the Lorentz 
algebras presented in Eq.~(\ref{cartangrasscliff}).}.

Let us use the technique which makes ''basis vectors''  products of nilpotents and projectors~\cite{norma93,norma95,nh02,nh03} which are eigenvectors of the 
(chosen) Cartan subalgebra members, Eq.~(\ref{cartangrasscliff}), of the Lorentz 
algebra in the space of $\gamma^{a}$'s, either  in the case of the Clifford odd 
or in the case of the Clifford even products of  $\gamma^{a}$'s .

There  are $\frac{d}{2}$ members of the Cartan subalgebra, 
Eq.~(\ref{cartangrasscliff}), in  even dimensional spaces.

One finds for any of the $\frac{d}{2}$ Cartan subalgebra member, $S^{ab}$ or 
$\tilde{S}^{ab}$, both applying on a nilpotent  $\stackrel{ab}{(k)}$ or on projector
$\stackrel{ab}{[k]}$
$$\stackrel{ab}{(k)}:=\frac{1}{2}(\gamma^a + 
\frac{\eta^{aa}}{ik} \gamma^b)\,, \;\;\; (\stackrel{ab}{(k)})^2=0,$$ 
 $$\stackrel{ab}{[k]}:=
\frac{1}{2}(1+ \frac{i}{k} \gamma^a \gamma^b)\,, \;\;\;(\stackrel{ab}{[k]})^2=
\stackrel{ab}{[k]}$$ 
the relations
\begin{eqnarray}
\label{signature0}
S^{ab} \,\stackrel{ab}{(k)} = \frac{k}{2}  \,\stackrel{ab}{(k)}\,,\quad && \quad
\tilde{S}^{ab}\,\stackrel{ab}{(k)} = \frac{k}{2}  \,\stackrel{ab}{(k)}\,,\nonumber\\
S^{ab}\,\stackrel{ab}{[k]} =  \frac{k}{2}  \,\stackrel{ab}{[k]}\,,\quad && \quad 
\tilde{S}^{ab} \,\stackrel{ab}{[k]} = - \frac{k}{2}  \,\,\stackrel{ab}{[k]}\,,
\end{eqnarray}
with  $k^2=\eta^{aa} \eta^{bb}$, demonstrating that the eigenvalues of 
$S^{ab}$ on nilpotents and projectors expressed with $\gamma^a$'s differ from the 
eigenvalues of $\tilde{S}^{ab}$ on  nilpotents and projectors expressed with 
$\gamma^a$'s, so that $\tilde{S}^{ab}$ can be used to equip each irreducible 
representation of $S^{ab}$ with the ''family'' quantum number.~\footnote{
 The reader can find the proof of Eq.~(\ref{signature0})  in Ref.~\cite{nh2021RPPNP}, 
 App.~(I).}

We define in even $d$ the ''basis vectors'' as algebraic, $*_A$, products of nilpotents
and projectors so that each product is eigenvector of all $\frac{d}{2}$ Cartan 
subalgebra members.

We recognize in advance that the superposition of an odd products of 
$\gamma^{a}$'s, that is the Clifford odd ''basis vectors'', must include an odd 
number of nilpotents, at least one, while the superposition of an even products 
of $\gamma^{a}$''s, that is Clifford even ''basis vectors'', must include an even 
number of nilpotents or only projectors. 

To define the Clifford odd ''basis vectors'', we shall see that they have properties 
appropriate to describe the internal space of the second quantized fermion fields,
and the Clifford even ''basis vectors'', we shall see that they have properties 
appropriate to describe the internal space of the second quantized boson fields, we 
need to know the relations for nilpotents and projectors
\begin{eqnarray}
\label{graficcliff}
\stackrel{ab}{(k)}:&=& 
\frac{1}{2}(\gamma^a + \frac{\eta^{aa}}{ik} \gamma^b)\,,\quad 
\stackrel{ab}{[k]}:=\frac{1}{2}(1+ \frac{i}{k} \gamma^a \gamma^b)\,,\nonumber\\
\stackrel{ab}{\tilde{(k)}}:&=& 
\frac{1}{2}(\tilde{\gamma}^a + \frac{\eta^{aa}}{ik} \tilde{\gamma}^b)\,,\quad 
\stackrel{ab}{\tilde{[k]}}:
\frac{1}{2}(1+ \frac{i}{k} \tilde{\gamma}^a \tilde{\gamma}^b)\,, 
\end{eqnarray}
which can be derived after taking into account Eq.~(\ref{gammatildeantiher})
\begin{small}
\begin{eqnarray}
%
\gamma^a \stackrel{ab}{(k)}&=& \eta^{aa}\stackrel{ab}{[-k]},\; \quad
\gamma^b \stackrel{ab}{(k)}= -ik \stackrel{ab}{[-k]}, \; \quad 
\gamma^a \stackrel{ab}{[k]}= \stackrel{ab}{(-k)},\;\quad \;\;
\gamma^b \stackrel{ab}{[k]}= -ik \eta^{aa} \stackrel{ab}{(-k)}\,,\nonumber\\
\tilde{\gamma^a} \stackrel{ab}{(k)} &=& - i\eta^{aa}\stackrel{ab}{[k]},\quad
\tilde{\gamma^b} \stackrel{ab}{(k)} =  - k \stackrel{ab}{[k]}, \;\qquad  \,
\tilde{\gamma^a} \stackrel{ab}{[k]} =  \;\;i\stackrel{ab}{(k)},\; \quad
\tilde{\gamma^b} \stackrel{ab}{[k]} =  -k \eta^{aa} \stackrel{ab}{(k)}\,, 
\nonumber\\
\stackrel{ab}{(k)}^{\dagger} &=& \eta^{aa}\stackrel{ab}{(-k)}\,,\quad 
(\stackrel{ab}{(k)})^2 =0\,, \quad \stackrel{ab}{(k)}\stackrel{ab}{(-k)}
=\eta^{aa}\stackrel{ab}{[k]}\,,\nonumber\\
\stackrel{ab}{[k]}^{\dagger} &=& \,\stackrel{ab}{[k]}\,, \quad \quad \quad \quad
(\stackrel{ab}{[k]})^2 = \stackrel{ab}{[k]}\,, 
\quad \stackrel{ab}{[k]}\stackrel{ab}{[-k]}=0\,,
\nonumber\\
\stackrel{ab}{(k)}\stackrel{ab}{[k]}& =& 0\,,\qquad \qquad \qquad 
\stackrel{ab}{[k]}\stackrel{ab}{(k)}=  \stackrel{ab}{(k)}\,, \quad \quad \quad
  \stackrel{ab}{(k)}\stackrel{ab}{[-k]} =  \stackrel{ab}{(k)}\,,
\quad \, \stackrel{ab}{[k]}\stackrel{ab}{(-k)} =0\,,
\nonumber\\
%
\stackrel{ab}{\tilde{(k)}}^{\dagger} &=& \eta^{aa}\stackrel{ab}{\tilde{(-k)}}\,,\quad
(\stackrel{ab}{\tilde{(k)}})^2=0\,, \quad \stackrel{ab}{\tilde{(k)}}\stackrel{ab}{\tilde{(-k)}}
=\eta^{aa}\stackrel{ab}{\tilde{[k]}}\,,\nonumber\\
\stackrel{ab}{\tilde{[k]}}^{\dagger} &=& \,\stackrel{ab}{\tilde{[k]}}\,,
\quad \quad \quad \quad
(\stackrel{ab}{\tilde{[k]}})^2=\stackrel{ab}{\tilde{[k]}}\,,
\quad \stackrel{ab}{\tilde{[k]}}\stackrel{ab}{\tilde{[-k]}}=0\,,\nonumber\\
\stackrel{ab}{\tilde{(k)}}\stackrel{ab}{\tilde{[k]}}& =& 0\,,\qquad \qquad \qquad 
\stackrel{ab}{\tilde{[k]}}\stackrel{ab}{\tilde{(k)}}=  \stackrel{ab}{\tilde{(k)}}\,, 
\quad \quad \quad
  \stackrel{ab}{\tilde{(k)}}\stackrel{ab}{\tilde{[-k]}} =  \stackrel{ab}{\tilde{(k)}}\,,
\quad \, \stackrel{ab}{\tilde{[k]}}\stackrel{ab}{\tilde{(-k)}} =0\,.
\label{graficcliff1}
\end{eqnarray}
\end{small}

 Looking at relations in Eq.~(\ref{graficcliff1}) it is obvious that the properties of the
 ''basis vectors'' which include odd number of nilpotents differer essentially from the
 ''basis vectors'' which include even number of nilpotents.
 
 One namely recognizes:\\
 {\bf i.} Since the Hermitian conjugated partner of a nilpotent 
 $\stackrel{ab}{(k)}^{\dagger}$ is $\eta^{aa}\stackrel{ab}{(-k)}$ and since 
 neither  $S^{ab}$ nor $\tilde{S}^{ab}$ nor both can transform odd products of 
 nilpotents  to belong to one of the $2^{\frac{d}{2}-1}$ members of one of
$2^{\frac{d}{2}-1}$ irreducible representations (families), 
the  Hermitian conjugated partners of the Clifford odd ''basis vectors'' must belong
to a different group of $2^{\frac{d}{2}-1}$ members of $2^{\frac{d}{2}-1}$ 
families.\\
\vspace{1mm}
\noindent
Since $S^{ac}$ transforms  $\stackrel{ab}{(k)} *_A  \stackrel{cd}{(k')}$ into
 $\stackrel{ab}{[-k]} *_A  \stackrel{cd}{[-k']}$, while $\tilde{S}^{ab}$ transforms
 $\stackrel{ab}{[-k]} *_A  \stackrel{cd}{[-k']}$ into $\stackrel{ab}{(-k)} *_A  
 \stackrel{cd}{(-k')}$ it is obvious that the Hermitian conjugated partners 
 of the Clifford odd ''basis vectors'' must belong to the same group  of 
 $2^{\frac{d}{2}-1}\times 2^{\frac{d}{2}-1}$ members. Projectors are 
 self adjoint. \\
 {\bf ii.} Since an odd products of $\gamma^{a}$'s anticommute with another group
 of an odd product of $\gamma^{a}$, the Clifford odd ''basis vectors'' anticommute,
 manifesting in a tensor product with the basis in ordinary space together  with the
 corresponding Hermitian conjugated partners properties of the anticommutation 
 relations postulated by Dirac for the second quantized fermion fields.\\ 
 \vspace{1mm}
\noindent
 The Clifford even  ''basis vectors''  correspondingly fulfil the commutation relations
 for the second quantized boson fields.\\
 {\bf iii.} The Clifford odd  ''basis vectors''  have all the eigenvalues of the Cartan 
 subalgebra members equal to either $\pm \frac{1}{2}$ or to  $\pm \frac{i}{2}$.\\
 \vspace{1mm}
\noindent
 The Clifford even  ''basis vectors''  have all the eigenvalues of the Cartan 
 subalgebra members ${\bf {\cal S}}^{ab}$ equal to either $\pm 1$ and zero
 or to $\pm i$ and zero.\\

Let us define odd an even ''basis vectors'' as products of nilpotents and projectors
in even dimensional spaces.

\vspace{3mm}

{\bf a.} $\;\;$ {\it Clifford odd ''basis vectors''}

\vspace{2mm}

The Clifford odd  "basis vectors'' must be products of an odd number of nilpotents and 
the rest, up to $\frac{d}{2}$,  of projectors,  each nilpotent and projector must be the ''eigenstates'' of one of the members of the Cartan subalgebra, Eq.~(\ref{cartangrasscliff}), correspondingly are the "basis vectors'' eigenstates of all 
the members of the Lorentz algebras: $S^{ab}$'s determine $2^{\frac{d}{2}-1}$ 
members of one family, $\tilde{S}^{ab}$'s transform  each member of one family to 
the same same member of the rest of $2^{\frac{d}{2}-1}$ families. 

Let us name the Clifford odd ''basis vectors'' $\hat{b}^{m \dagger}_{f}$, where $m$ 
determines membership  of 'basis vectors'' in any  family and $f$ determines a particular 
family.  The Hermitian conjugated partner of $\hat{b}^{m \dagger}_{f}$ is named by 
$\hat{b}^{m}_{f}=(\hat{b}^{m \dagger}_{f})^{\dagger}$.

Let us start in  $d=2(2n+1)$ with the ''basis vector'' $\hat{b}^{1 \dagger}_{1}$ 
which is the product of only nilpotents, all the rest members belonging to the $f=1$ 
family follow by the application of $S^{01}$, $S^{03}$, $ \dots, S^{0d}, S^{15}$,
$\dots, S^{1d}, S^{5 d}\dots, S^{d-2\, d}$. The algebraic product mark $*_{A}$ is 
skipped.
\begin{small}
\begin{eqnarray}
\label{allcartaneigenvec}
&& d=2(2n+1)\, ,\nonumber\\
&& \hat{b}^{1 \dagger}_{1}=\stackrel{03}{(+i)}\stackrel{12}{(+)} \stackrel{56}{(+)}
\cdots \stackrel{d-1 \, d}{(+)}\,,\nonumber\\
&&\hat{b}^{2 \dagger}_{1} = \stackrel{03}{[-i]} \stackrel{12}{[-]} 
\stackrel{56}{(+)} \cdots \stackrel{d-1 \, d}{(+)}\,,\nonumber\\
&& \cdots\nonumber\\
&&\hat{b}^{2^{\frac{d}{2}-1} \dagger}_{1} = \stackrel{03}{[-i]} \stackrel{12}{[-]} 
\stackrel{56}{(+)} \dots \stackrel{d-3\,d-2}{[-]}\;\stackrel{d-1\,d}{[-]}\,, \nonumber\\
&& \cdots\,.
\end{eqnarray}
\end{small}

The Hermitian conjugated partners of the Clifford odd ''basis vector'' 
$\hat{b}^{m \dagger}_{1}$, presented in Eq.~(\ref{allcartaneigenvec}),  are
\begin{small}
\begin{eqnarray}
\label{allcartaneigenvecher}
&& d=2(2n+1)\, ,\nonumber\\
&& \hat{b}^{1}_{1}=\stackrel{03}{(-i)}\stackrel{12}{(-)}
\cdots \stackrel{d-1 \, d}{(-)}\,,\nonumber\\
&&\hat{b}^{2 }_{1} = \stackrel{03}{[-i]} \stackrel{12}{[-]} 
\stackrel{56}{(-)} \cdots \stackrel{d-1 \, d}{(-)}\,,\nonumber\\
&& \cdots\nonumber\\
&&\hat{b}^{2^{\frac{d}{2}-1} \dagger}_{1} = \stackrel{03}{[-i]} \stackrel{12}{[-]} 
\stackrel{56}{(-)} \stackrel{78}{[-]} \dots \stackrel{d-3\,d-2}{[-]}\;\stackrel{d-1\,d}{[-]}\,, \nonumber\\
&& \cdots\,.
\end{eqnarray}
\end{small}

In $d=4n$ the choice of the starting ''basis vector''with maximal number of nilpotents
must have one projector
\begin{small}
\begin{eqnarray}
\label{allcartaneigenvec4n}
&& d=4n\, ,\nonumber\\
&& \hat{b}^{1 \dagger}_{1}=\stackrel{03}{(+i)}\stackrel{12}{(+)}
\cdots \stackrel{d-1 \, d}{[+]}\,,\nonumber\\
&&\hat{b}^{2 \dagger}_{1} = \stackrel{03}{[-i]} \stackrel{12}{[-]} 
\stackrel{56}{(+)} \cdots \stackrel{d-1 \, d}{[+]}\,,\nonumber\\
&& \cdots\nonumber\\
&&\hat{b}^{2^{\frac{d}{2}-1} \dagger}_{1} = \stackrel{03}{[-i]} \stackrel{12}{[-]} 
\stackrel{56}{(+)} \dots \stackrel{d-3\,d-2}{[-]}\;\stackrel{d-1\,d}{[+]}\,, \nonumber\\
&& \dots\,.
\end{eqnarray}
\end{small}
The Hermitian conjugated partners of the Clifford odd ''basis vector'' 
$\hat{b}^{m \dagger}_{1}$, presented in Eq.~(\ref{allcartaneigenvec4n}), follow
if all nilpotents $\stackrel{ab}{(k)}$  are transformed into $\eta^{aa}
\stackrel{ab}{(-k)}$.

For either $d=2(2n+1)$ or for $d=4n$ all the $2^{\frac{d}{2}-1}$ families follow by 
applying $\tilde{S}^{ab}$'s on all the members of the starting family. (Or one can find 
the starting $ \hat{b}^{1}_{f}$ for all families $f$ and then generate all the members  
$\hat{b}^{m}_{f}$ from  $\hat{b}^{1}_{f}$ by the application of $\tilde{S}^{ab}$
on the starting member.)

It is not difficult to see that all the ''basis vectors''  within any family as well as  the 
''basis vectors'' among families are orthogonal, that is their algebraic product is zero, 
and the same is true for the Hermitian conjugated partners, what can be proved by
the  algebraic multiplication using  Eq.(\ref{graficcliff1}). 

\begin{eqnarray}
\hat{b}^{m \dagger}_f *_{A} \hat{b}^{m `\dagger }_{f `}&=& 0\,, 
\quad \hat{b}^{m}_f *_{A} \hat{b}^{m `}_{f `}= 0\,, \quad \forall m,m',f,f `\,. 
\label{orthogonalodd}
\end{eqnarray}

If we require that each family of ''basis vectors'', determined by nilpotents and 
projectors described by $\gamma^{a}$'s, carry the family quantum number 
determined by $\tilde{S}^{ab}$ and define the vacuum state on which ''basis 
vectors''  apply as
\begin{eqnarray}
\label{vaccliffodd}
|\psi_{oc}>= \sum_{f=1}^{2^{\frac{d}{2}-1}}\,\hat{b}^{m}_{f}{}_{*_A}
\hat{b}^{m \dagger}_{f} \,|\,1\,>\,,
\end{eqnarray}
it follows that the Clifford odd ''basis vectors'' obey the relations
\begin{eqnarray}
\label{almostDirac}
\hat{b}^{m}_{f} {}_{*_{A}}|\psi_{oc}>&=& 0.\, |\psi_{oc}>\,,\nonumber\\
\hat{b}^{m \dagger}_{f}{}_{*_{A}}|\psi_{oc}>&=&  |\psi^m_{f}>\,,\nonumber\\
\{\hat{b}^{m}_{f}, \hat{b}^{m'}_{f `}\}_{*_{A}+}|\psi_{oc}>&=&
 0.\,|\psi_{oc}>\,, \nonumber\\
\{\hat{b}^{m \dagger}_{f}, \hat{b}^{m' \dagger}_{f  `}\}_{*_{A}+}|\psi_{oc}>
&=& 0. \,|\psi_{oc}>\,,\nonumber\\
\{\hat{b}^{m}_{f}, \hat{b}^{m' \dagger}_{f }\}_{*_{A}+}|\psi_{oc}>
&=& \delta^{m m'} \,\delta_{f f `}|\psi_{oc}>\,,
\end{eqnarray}
while  the normalization 
$<\psi_{oc}| \hat{b}^{m' \dagger}_{f'}\, *_{A}\,\hat{b}^{m \dagger}_{f}
*_{A}|\psi_{oc}> = \delta^{m m'} \delta_{f f'}\,$  is used and the anticommutation
relation mean $\{\hat{b}^{m \dagger}_{f}, \hat{b}^{m' \dagger}_{f  `}\}_{*_{A}+}=$
$\hat{b}^{m \dagger}_{f} \,*_A\, \hat{b}^{m' \dagger}_{f  `}+
\hat{b}^{m' \dagger}_{f `} \,*_A \,\hat{b}^{m \dagger}_{f }$.

If we write the creation and annihilation operators as the tensor, $*_{T}$,  products
of ''basis vectors'' and the basis in ordinary space, the creation and annihilation 
operators fulfil the Dirac's anticommutation postulates since the ''basis vectors'' transfer 
their anticommutativity to creation and annihilation operators. It turns out that not only
the Clifford odd ''basis vectors'' offer the description of the internal space of fermions, 
they  offer the explanation for the second quantization postulates for fermions as well. 

Table~\ref{Table Clifffourplet.},  presented in Subsect.~\ref{cliffordoddevenbasis5+1},
illustrates the properties of the Clifford odd ''basis vectors'' on the case of $d=(5+1)$.

\vspace{3mm}

{\bf b.} $\;\;$ {\it Clifford even ''basis vectors''}

\vspace{2mm}

The Clifford even  "basis vectors'' must be products of an even number of nilpotents and 
the rest, up to $\frac{d}{2}$,  of projectors,  each nilpotent and projector in a product 
must be the ''eigenstate'' of one of the members of the Cartan subalgebra, Eq.~(\ref{cartangrasscliff}), correspondingly are the "basis vectors'' eigenstates of all 
the members of the Lorentz algebra: $S^{ab}$'s and $\tilde{S}^{ab}$'s generate 
from the starting "basis vector'' all the $2^{\frac{d}{2}-1} \times$ $2^{\frac{d}{2}-1}$ 
members of one group which includes as well the Hermitian conjugated partners of any 
member.  $2^{\frac{d}{2}-1}$ members of the group are products of projectors only.
They are self adjoint.

There are two groups with $2^{\frac{d}{2}-1}\times $$2^{\frac{d}{2}-1}$ 
members. The members of one group are not connected with the members of another
group by either by $S^{ab}$'s or $\tilde{S}^{ab}$'s or both.

Let us name the Clifford even ''basis vectors'' ${}^{i}\hat{\cal A}^{m \dagger}_{f}$, 
where $i=(I,II)$ denotes that there are two groups of Clifford even ''basis vectors'', while
$m$  and $f$ determine membership  of 'basis vectors'' in any of the two groups $I$
or $II$. Let me repeat that the Hermitian conjugated partner of any ''basis vector''
appears either in the case of ${}^{I}\hat{\cal A}^{m \dagger}_{f}$ or in the case of 
${}^{II}\hat{\cal A}^{m \dagger}_{f}$ within the same group.

Let us write down the Clifford even ''basis vectors'' as a product of an even number of 
nilpotents and the rest of projectors, so that the Clifford even ''basis vectors'' are 
eigenvectors of all the Cartan subalgebra members, and let us name them as follows
\begin{eqnarray}
\label{allcartaneigenvecevenI} 
d&=&2(2n+1)\nonumber\\
{}^I\hat{{\cal A}}^{1 \dagger}_{1}=\stackrel{03}{(+i)}\stackrel{12}{(+)}\cdots 
\stackrel{d-1 \, d}{[+]}\,,\qquad &&
{}^{II}\hat{{\cal A}}^{1 \dagger}_{1}=\stackrel{03}{(-i)}\stackrel{12}{(+)}\cdots 
\stackrel{d-1 \, d}{[+]}\,,\nonumber\\
{}^I\hat{{\cal A}}^{2 \dagger}_{1}=\stackrel{03}{[-i]}\stackrel{12}{[-]} 
\stackrel{56}{(+)} \cdots \stackrel{d-1 \, d}{[+]}\,, \qquad  && 
{}^{II}\hat{{\cal A}}^{2 \dagger}_{1}=\stackrel{03}{[+i]}\stackrel{12}{[-]} 
\stackrel{56}{(+)} \cdots \stackrel{d-1 \, d}{[+]}\,,
\nonumber\\ 
{}^I\hat{{\cal A}}^{3 \dagger}_{1}=\stackrel{03}{(+i)} \stackrel{12}{(+)} 
\stackrel{56}{(+)} \cdots \stackrel{d-3\,d-2}{[-]}\;\stackrel{d-1\,d}{(-)}\,, \qquad &&
{}^{II}\hat{{\cal A}}^{3 \dagger}_{1}=\stackrel{03}{(-i)} \stackrel{12}{(+)} 
\stackrel{56}{(+)} \cdots \stackrel{d-3\,d-2}{[-]}\;\stackrel{d-1\,d}{(-)}\,,  \nonumber\\
\dots \qquad && \dots \nonumber\\
d&=&4n\nonumber\\
{}^I\hat{{\cal A}}^{1 \dagger}_{1}=\stackrel{03}{(+i)}\stackrel{12}{(+)}\cdots 
\stackrel{d-1 \, d}{(+)}\,,\qquad &&
{}^{II}\hat{{\cal A}}^{1 \dagger}_{1}=\stackrel{03}{(-i)}\stackrel{12}{(+)}\cdots 
\stackrel{d-1 \, d}{(+)}\,,
\nonumber\\
{}^I\hat{{\cal A}}^{2 \dagger}_{1}= \stackrel{03}{[-i]}\stackrel{12}{[-i]} 
\stackrel{56}{(+)} \cdots \stackrel{d-1 \, d}{(+)}\,, \qquad &&
{}^{II}\hat{{\cal A}}^{2 \dagger}_{1}= \stackrel{03}{[+i]}\stackrel{12}{[-i]} 
\stackrel{56}{(+)} \cdots \stackrel{d-1 \, d}{(+)}\,,  \nonumber\\ 
{}^I\hat{{\cal A}}^{3 \dagger}_{1}=\stackrel{03}{(+i)} \stackrel{12}{(+)} 
\stackrel{56}{(+)} \cdots \stackrel{d-3\,d-2}{[-]}\;\stackrel{d-1\,d}{[-]}\,, \qquad &&
{}^{II}\hat{{\cal A}}^{3 \dagger}_{1}=\stackrel{03}{(-i)} \stackrel{12}{(+)} 
\stackrel{56}{(+)} \cdots \stackrel{d-3\,d-2}{[-]}\;\stackrel{d-1\,d}{[-]}\,\nonumber\\
\dots \qquad && \dots 
\end{eqnarray}
%
There are $2^{\frac{d}{2}-1}\times 2^{\frac{d}{2}-1}$ Clifford  even ''basis vectors'' of
the kind ${}^{I}{\hat{\cal A}}^{m \dagger}_{f}$ and  there are $2^{\frac{d}{2}-1}$
$\times 2^{\frac{d}{2}-1}$ Clifford  even ''basis vectors'' of the kind
${}^{II}{\hat{\cal A}}^{m \dagger}_{f}$.

Table~\ref{Table Clifffourplet.}, presented in Subsect.~\ref{cliffordoddevenbasis5+1}, 
illustrates properties of the Clifford odd and Clifford even ''basis vectors'' on the 
case of $d=(5+1)$. Looking at this particular case it is easy to evaluate properties of 
either even or odd ''basis vectors''.  I shall present here the general results which 
follow after careful inspection of properties of both kinds of ''basis vectors''.

The Clifford even ''basis vectors''  belonging two two different groups are orthogonal due 
to the fact that they differ in the sign of one nilpotent or one projectors, or the algebraic 
products of members of one group give zero according to Eq.~(\ref{graficcliff1}).
\begin{eqnarray}
\label{AIAIIorth}
{}^{I}{\hat{\cal A}}^{m \dagger}_{f} *_A {}^{II}{\hat{\cal A}}^{m \dagger}_{f} 
&=&0={}^{II}{\hat{\cal A}}^{m \dagger}_{f} *_A 
{}^{I}{\hat{\cal A}}^{m \dagger}_{f}\,.
\end{eqnarray}
The members of each of this two groups have the property 
\begin{eqnarray}
\label{ruleAAI}
{}^{I,II}{\hat{\cal A}}^{m \dagger}_{f} \,*_A\, {}^{I,II}{\hat{\cal A}}^{m' \dagger}_{f `}
\rightarrow \left \{ \begin{array} {r} {}^{I,II}{\hat{\cal A}}^{m \dagger}_{f `}\,, 
{\rm only\; one\; for }\;
\forall f `\,,\\
{\rm or \,zero}\,.
\end{array} \right.
\end{eqnarray}
Two ''basis vectors'' ${}^{I}{\hat{\cal A}}^{m \dagger}_{f}$  and 
${}^{I}{\hat{\cal A}}^{m' \dagger}_{f '}$, the algebraic product, $*_{A}$, of which 
 gives nonzero contribution, ''scatter'' into the third one  
 ${}^{I}{\hat{\cal A}}^{m \dagger}_{f `}$. The same is true also for the ''basis vectors'' 
 ${}^{II}{\hat{\cal A}}^{m \dagger}_{f}$.

Let us write the commutation relations for Clifford even ''basis vectors'' taking into account Eq.~(\ref{ruleAAI}).\\

$\;\;$ {\bf i.} $\;\;$ In the case that ${}^{I}{\hat{\cal A}}^{m \dagger}_{f} \,*_{A} \, 
{}^{I}{\hat{\cal A}}^{m' \dagger}_{f `} \rightarrow $ 
$ {}^{I}{\hat{\cal A}}^{m \dagger}_{f `}$ and 
${}^{I}{\hat{\cal A}}^{m' \dagger}_{f `} \,*_{A} \,
{}^{I}{\hat{\cal A}}^{m \dagger}_{f}= 0$ it follows
\begin{eqnarray}
\label{ruleAAI1}
\{{}^{I}{\hat{\cal A}}^{m \dagger}_{f} \,, \, 
{}^{I}{\hat{\cal A}}^{m' \dagger}_{f `}\}_{*_A \,-}
\rightarrow \left \{ \begin{array} {r} {}^{I}{\hat{\cal A}}^{m \dagger}_{f `}\,, \;\;\;\;
({\rm if\; }\quad \;{}^{I}{\hat{\cal A}}^{m \dagger}_{f} \,*_{A} \, 
{}^{I}{\hat{\cal A}}^{m' \dagger}_{f `} 
\rightarrow  {}^{I}{\hat{\cal A}}^{m \dagger}_{f `}
\\
{\rm and }\;\;{}^{I}{\hat{\cal A}}^{m' \dagger}_{f `} \,*_{A} \,
{}^{I}{\hat{\cal A}}^{m \dagger}_{f}= 0)\,,
\end{array} \right.
\end{eqnarray}\\

 $\;\;$ {\bf ii.} $\;\;$ In  the case that 
 ${}^{I}{\hat{\cal A}}^{m \dagger}_{f} \,*_{A} \, 
{}^{I}{\hat{\cal A}}^{m' \dagger}_{f `} \rightarrow $ 
$ {}^{I}{\hat{\cal A}}^{m \dagger}_{f `}$ and 
${}^{I}{\hat{\cal A}}^{m' \dagger}_{f `} \,*_{A} \,
{}^{I}{\hat{\cal A}}^{m \dagger}_{f}\rightarrow $ 
${}^{I}{\hat{\cal A}}^{m' \dagger}_{f } \,$ it follows
\begin{eqnarray}
\label{ruleAAI2}
\{{}^{I}{\hat{\cal A}}^{m \dagger}_{f} \,, \, 
{}^{I}{\hat{\cal A}}^{m' \dagger}_{f `} \}_{*_A \,-}
\rightarrow \left \{ \begin{array} {r} {}^{I}{\hat{\cal A}}^{m \dagger}_{f `}
- {}^{I}{\hat{\cal A}}^{m' \dagger}_{f }\,, \;\;\;
({\rm if\; } \quad \;{}^{I}{\hat{\cal A}}^{m \dagger}_{f} \,*_{A} \, 
{}^{I}{\hat{\cal A}}^{m' \dagger}_{f `} 
\rightarrow  {}^{I}{\hat{\cal A}}^{m \dagger}_{f `}
\\
{\rm and }\;\;{}^{I}{\hat{\cal A}}^{m' \dagger}_{f `} \,*_{A} \,
{}^{I}{\hat{\cal A}}^{m \dagger}_{f}\rightarrow 
{}^{I}{\hat{\cal A}}^{m' \dagger}_{f})\,,
\end{array} \right.
\end{eqnarray}\\
%

 $\;\;$ {\bf iii.} $\;\;$ In all other cases we have
\begin{eqnarray}
\label{ruleAAI3}
\{{}^{I}{\hat{\cal A}}^{m \dagger}_{f} \,,\,
{}^{I}{\hat{\cal A}}^{m' \dagger}_{f `}\}_{*_A \,-} =0\;.
\end{eqnarray}
$\{{}^{I}{\hat{\cal A}}^{m \dagger}_{f} \,, \, 
{}^{I}{\hat{\cal A}}^{m' \dagger}_{f `} \}_{*_A \,-}$ means
${}^{I}{\hat{\cal A}}^{m \dagger}_{f} \, *_A \, 
{}^{I}{\hat{\cal A}}^{m' \dagger}_{f `}  - {}^{I}{\hat{\cal A}}^{m' \dagger}_{f `} \, *_A \, 
{}^{I}{\hat{\cal A}}^{m \dagger}_{f }$.\\

\vspace{2mm}
 
It remains to evaluate the algebraic application, $*_{A}$, of the Clifford even ''basis vectors'' 
${}^{I}{\hat{\cal A}}^{m \dagger}_{f }$ on the Clifford odd ''basis vectors'' 
$ \hat{b}^{m' \dagger}_{f `} $. One finds  
\begin{eqnarray}
\label{calAb1234gen}
{}^{I}{\hat{\cal A}}^{m \dagger}_{f `} \,*_A \, \hat{b}^{m' \dagger }_{f}
\rightarrow \left \{ \begin{array} {r} \hat{b }^{m \dagger}_{f }\,, \\
{\rm or \,zero}\,.
\end{array} \right.
\end{eqnarray}
For each ${}^{I}{\hat{\cal A}}^{m \dagger}_{f}$  there are among 
$2^{\frac{d}{2}-1}\times 2^{\frac{d}{2}-1}$ members of the Clifford odd 
''basis vectors'' (describing the internal space of fermion fields) 
$2^{\frac{d}{2}-1}$ members, $\hat{b}^{m' \dagger}_{f `}$, fulfilling the
relation of Eq.~(\ref{calAb1234gen}). All the rest ($2^{\frac{d}{2}-1}\times 
(2^{\frac{d}{2}-1}-1)$, give zero contributions.\\

Eq.~(\ref{calAb1234gen}) clearly demonstrates that 
${}^{I}{\hat{\cal A}}^{m \dagger}_{f}$ transforms the Clifford odd ''basis vector''
in general into another Clifford  odd   ''basis vector'', transfering to the   Clifford odd 
''basis vector''   an integer spin.

We can obviously conclude that the Clifford even ''basis vectors'' offer the description 
of the gauge fields to the corresponding fermion fields.

While  the Clifford odd ''basis vectors'' offer the description of the internal space of 
the second quantized anticommuting fermion fields, appearing in families,  the 
Clifford even ''basis vectors'' offer the description of the internal space of 
the second quantized commuting boson fields, having no families and manifesting 
as the gauge fields of the corresponding fermion fields.

\vspace{1mm}
\subsection{Example demonstrating properties of  Clifford odd and even
''basis vectors'' for $d=(5+1)$   }
`%
\label{cliffordoddevenbasis5+1}

\vspace{2mm}

Subsect.~\ref{cliffordoddevenbasis5+1} demonstrates properties of the Clifford 
odd and even ''basis vectors'' in a special case when $d=(5+1)$.

In Table~\ref{Table Clifffourplet.} the $64 \,(=2^{d=6})$ ''eigenvectors" of the Cartan 
subalgebra  members of the Lorentz algebra $S^{ab}$, Eq.~(\ref{cartangrasscliff}), 
are presented. The Clifford odd "basis vectors'', they appear in $4 \,(=2^{\frac{d=6}{2}-1})$ 
families, each family has $4$ members, are products of an odd number of nilpotents, 
that is either of three nilpotents or of one nilpotent. They appear  in 
Table~\ref{Table Clifffourplet.} in  the group  named $odd \,I \,\hat{b}^{m\dagger}_f$. 
Their Hermitian conjugated partners appear in the second group  named 
$odd \,II \,\hat{b}^m_f$. Within each of these two groups, the members are orthogonal, 
Eq.~(\ref{orthogonalodd}), which means that the algebraic product of 
$\hat{b}^{m\dagger} _f *_{A} \hat{b}^{m'\dagger} _{f `} =0$ for all $(m,m', f, f `)$,
what can be checked by using relations in Eq.~(\ref{graficcliff1}). Equivalently,  the
algebraic products of their Hermitian conjugated partners are  also orthogonal among 
themselves.  
The ''basis vectors'' and their Hermitian  conjugated partners are normalized as follows
\begin{eqnarray}
< \psi_{oc}| \hat{b}^{m} _f *_{A} \hat{b}^{m' \dagger} _{f `} |\psi_{oc} >=
\delta^{m m'} \delta_{f f `}\,, 
\label{Cliffnormalizationodd}
\end{eqnarray}
since the vacuum state $|\psi_{oc} >= \frac{1}{\sqrt{2^{\frac{d=6}{2}-1}}}$
$(\stackrel{03}{[-i]} \stackrel{12}{[-]}\stackrel{56}{[-]}+ \stackrel{03}{[-i]} 
\stackrel{12}{[+]}\stackrel{56}{[+]} + \stackrel{03}{[+i]} \stackrel{12}{[-]}
\stackrel{56}{[+]}+ \stackrel{03}{[+i]} \stackrel{12}{[+]}\stackrel{56}{[-]})$
is normalized to one: $< \psi_{oc}|\psi_{oc} >=1$.

The longer overview of the properties of the Clifford odd ''basis vectors'' and their 
Hermitian conjugated partners for the case $d=(5+1)$  can be found in 
Ref.~\cite{nh2021RPPNP}.

The Clifford even "basis vectors'' are products of an even number of nilpotents, of 
either two  or none in this case. They are presented  in Table~\ref{Table Clifffourplet.} 
in two groups, each with $16\, (=2^{\frac{d=6}{2}-1}\times 2^{\frac{d=6}{2}-1})$  
members, as $even \, I \, {\bf {\cal A}}^{m \dagger}_{f}$ and 
$even \, II \, {\bf {\cal A}}^{m \dagger}_{f}$. One can easily check, using 
Eq.~(\ref{graficcliff1}), that the algebraic product 
${}^{I} {\bf {\cal A}}^{m \dagger}_{f} *_{A}$
${}^{II} {\bf {\cal A}}^{m' \dagger}_{f `} =0, \forall\, (m,m',f. f `)$, 
Eq.~(\ref{AIAIIorth}).
The longer overview of the Clifford even ''basis vectors'' and their Hermitian 
conjugated partners for the case $d=(5+1)$- can be found in 
Ref.~\cite{n2021SQ}.

\begin{table*}
\begin{small}
\caption{\label{Table Clifffourplet.}  $2^d=64$ "eigenvectors" of the Cartan subalgebra
of the Clifford  odd and even algebras --- the superposition of odd and 
even products of $\gamma^{a}$'s --- in $d=(5+1)$-dimensional space are presented, 
divided into four groups. The first group, $odd \,I$, is chosen to represent "basis vectors", 
named ${\hat b}^{m \dagger}_f$, appearing in $2^{\frac{d}{2}-1}=4$ 
"families"  ($f=1,2,3,4$), each ''family'' with $2^{\frac{d}{2}-1}=4$  
''family'' members ($m=1,2,3,4$). 
The second group, $odd\,II$, contains Hermitian conjugated partners of the first 
group for each  family separately, ${\hat b}^{m}_f=$ 
$({\hat b}^{m \dagger}_f)^{\dagger}$. Either $odd \,I$ or $odd \,II$ are products
of an odd number of nilpotents, the rest are projectors.
The "family" quantum  numbers of ${\hat b}^{m \dagger}_f$, that is the eigenvalues of 
$(\tilde{S}^{03}, \tilde{S}^{12},\tilde{S}^{56})$,  are for the first {\it odd I } 
group written   above each "family", the quantum 
numbers of the members  $(S^{03}, S^{12}, S^{56})$ are 
 written in the last three columns. 
For the Hermitian conjugated  partners of {\it odd I}, presented in the group {\it odd II},
the quantum numbers $(S^{03}, S^{12}, S^{56})$ are presented above each group of the
Hermitian conjugated partners, the last three columns 
tell eigenvalues of $(\tilde{S}^{03}, \tilde{S}^{12},\tilde{S}^{56})$.
The  two groups with the even number of $\gamma^a$'s, {\it even \,I} and {\it even \,II}, 
each has their Hermitian conjugated partners within its own group,
have the  quantum  numbers  $f$, that is the eigenvalues of 
$(\tilde{S}^{03}, \tilde{S}^{12},\tilde{S}^{56})$, written   above column of 
four members, the quantum numbers of the members,  $(S^{03}, S^{12}, S^{56})$, are
 written in the last three columns.
 \vspace{2mm}}
 \end{small}
\begin{tiny}
\begin{center}
  \begin{tabular}{|c|c|c|c|c|c|r|r|r|}
\hline
$ $&$$&$ $&$ $&$ $&&$$&$$&$$\\
$''basis\, vectors'' $&$m$&$ f=1$&$ f=2 $&$ f=3 $&
$ f=4 $&$$&$$&$$\\ 
$(\tilde{S}^{03}, \tilde{S}^{12}, \tilde{S}^{56})$&$\rightarrow$&$(\frac{i}{2},- \frac{1}{2},-\frac{1}{2})$&$(-\frac{i}{2},-\frac{1}{2},\frac{1}{2})$&
$(-\frac{i}{2},\frac{1}{2},-\frac{1}{2})$&$(\frac{i}{2},\frac{1}{2},\frac{1}{2})$&$S^{03}$
 &$S^{12}$&$S^{56}$\\ 
\hline
$ $&$$&$ $&$ $&$ $&&$$&$$&$$\\ 
$odd \,I\; {\hat b}^{m \dagger}_f$&$1$& 
$\stackrel{03}{(+i)}\stackrel{12}{[+]}\stackrel{56}{[+]}$&
                        $\stackrel{03}{[+i]}\stackrel{12}{[+]}\stackrel{56}{(+)}$ & 
                        $\stackrel{03}{[+i]}\stackrel{12}{(+)}\stackrel{56}{[+]}$ &  
                        $\stackrel{03}{(+i)}\stackrel{12}{(+)}\stackrel{56}{(+)}$ &
                        $\frac{i}{2}$&$\frac{1}{2}$&$\frac{1}{2}$\\ 
$$&$2$&    $[-i](-)[+] $ & $(-i)(-)(+) $ & $(-i)[-][+] $ & $[-i][-](+) $ &$-\frac{i}{2}$&
$-\frac{1}{2}$&$\frac{1}{2}$\\ 
$$&$3$&    $[-i] [+](-)$ & $(-i)[+][-] $ & $(-i)(+)(-) $ & $[-i](+)[-] $&$-\frac{i}{2}$&
$\frac{1}{2}$&$-\frac{1}{2}$\\ 
$$&$4$&    $(+i)(-)(-)$ & $[+i](-)[-] $ & $[+i][-](-) $ & $(+i)[-][-]$&$\frac{i}{2}$&
$-\frac{1}{2}$&$-\frac{1}{2}$\\ 
\hline
$ $&$$&$ $&$ $&$ $&&$$&$$&$$\\ 
$(S^{03}, S^{12}, S^{56})$&$\rightarrow$&$(-\frac{i}{2}, \frac{1}{2},\frac{1}{2})$&
$(\frac{i}{2},\frac{1}{2},-\frac{1}{2})$&
$(\frac{i}{2},- \frac{1}{2},\frac{1}{2})$&$(-\frac{i}{2},-\frac{1}{2},-\frac{1}{2})$&
$\tilde{S}^{03}$
&$\tilde{S}^{12}$&$\tilde{S}^{56}$\\ 
&&
$\stackrel{03}{\;\,}\;\;\,\stackrel{12}{\;\,}\;\;\,\stackrel{56}{\;\,}$&
$\stackrel{03}{\;\,}\;\;\,\stackrel{12}{\;\,}\;\;\,\stackrel{56}{\;\,}$&
$\stackrel{03}{\;\,}\;\;\,\stackrel{12}{\;\,}\;\;\,\stackrel{56}{\;\,}$&
$\stackrel{03}{\;\,}\;\;\,\stackrel{12}{\;\,}\;\;\,\stackrel{56}{\;\,}$&
&&\\
\hline
$ $&$$&$ $&$ $&$ $&&$$&$$&$$\\ 
$odd\,II\; {\hat b}^{m}_f$&$1$ &$(-i)[+][+]$ & $[+i][+](-)$ & $[+i](-)[+]$ & $(-i)(-)(-)$&
$-\frac{i}{2}$&$-\frac{1}{2}$&$-\frac{1}{2}$\\ 
$$&$2$&$[-i](+)[+]$ & $(+i)(+)(-)$ & $(+i)[-][+]$ & $[-i][-](-)$&
$\frac{i}{2}$&$\frac{1}{2}$&$-\frac{1}{2}$\\ 
$$&$3$&$[-i][+](+)$ & $(+i)[+][-]$ & $(+i)(-)(+)$ & $[-i](-)[-]$&
$\frac{i}{2}$&$-\frac{1}{2}$&$\frac{1}{2}$\\ 5 &$-1$&$-1$ \\
$$&$4$&$(-i)(+)(+)$ & $[+i](+)[-]$ & $[+i][-](+)$ & $(-i)[-][-]$&
$-\frac{i}{2}$&$\frac{1}{2}$&$\frac{1}{2}$\\ 
\hline
&&&&&&&&\\ 
\hline
$ $&$$&$ $&$ $&$ $&&$$&$$&$$\\ 
$(\tilde{S}^{03}, \tilde{S}^{12}, \tilde{S}^{56})$&$\rightarrow$&
$(-\frac{i}{2},\frac{1}{2},\frac{1}{2})$&$(\frac{i}{2},-\frac{1}{2},\frac{1}{2})$&
$(-\frac{i}{2},-\frac{1}{2},-\frac{1}{2})$&$(\frac{i}{2},\frac{1}{2},-\frac{1}{2})$&
$S^{03}$&$S^{12}$&$S^{56}$\\ 
&& 
$\stackrel{03}{\;\,}\;\;\,\stackrel{12}{\;\,}\;\;\,\stackrel{56}{\;\,}$&
$\stackrel{03}{\;\,}\;\;\,\stackrel{12}{\;\,}\;\;\,\stackrel{56}{\;\,}$&
$\stackrel{03}{\;\,}\;\;\,\stackrel{12}{\;\,}\;\;\,\stackrel{56}{\;\,}$&
$\stackrel{03}{\;\,}\;\;\,\stackrel{12}{\;\,}\;\;\,\stackrel{56}{\;\,}$&
&&\\ 
\hline
$ $&$$&$ $&$ $&$ $&&$$&$$&$$\\ 
$even\,I \; {}^{I}{\cal A}^{m}_f$&$1$&$[+i](+)(+) $ & $(+i)[+](+) $ & $[+i][+][+] $ & $(+i)(+)[+] $ &$\frac{i}{2}$&
$\frac{1}{2}$&$\frac{1}{2}$\\ 
$$&$2$&$(-i)[-](+) $ & $[-i](-)(+) $ & $(-i)(-)[+] $ & $[-i][-][+] $ &$-\frac{i}{2}$&
$-\frac{1}{2}$&$\frac{1}{2}$\\ 
$$&$3$&$(-i)(+)[-] $ & $[-i][+][-] $ & $(-i)[+](-) $ & $[-i](+)(-) $&$-\frac{i}{2}$&
$\frac{1}{2}$&$-\frac{1}{2}$\\ 
$$&$4$&$[+i][-][-] $ & $(+i)(-)[-] $ & $[+i](-)(-) $ & $(+i)[-](-) $&$\frac{i}{2}$&
$-\frac{1}{2}$&$-\frac{1}{2}$\\ 
\hline
$ $&$$&$ $&$ $&$ $&&$$&$$&$$\\ 
$(\tilde{S}^{03}, \tilde{S}^{12}, \tilde{S}^{56})$&$\rightarrow$&
$(\frac{i}{2},\frac{1}{2},\frac{1}{2})$&$(-\frac{i}{2},-\frac{1}{2},\frac{1}{2})$&
$(\frac{i}{2},-\frac{1}{2},-\frac{1}{2})$&$(-\frac{i}{2},\frac{1}{2},-\frac{1}{2})$&
$S^{03}$&$S^{12}$&$S^{56}$\\ 
&& 
$\stackrel{03}{\;\,}\;\;\,\stackrel{12}{\;\,}\;\;\,\stackrel{56}{\;\,}$&
$\stackrel{03}{\;\,}\;\;\,\stackrel{12}{\;\,}\;\;\,\stackrel{56}{\;\,}$&
$\stackrel{03}{\;\,}\;\;\,\stackrel{12}{\;\,}\;\;\,\stackrel{56}{\;\,}$&
$\stackrel{03}{\;\,}\;\;\,\stackrel{12}{\;\,}\;\;\,\stackrel{56}{\;\,}$&
&&\\ 
\hline
$ $&$$&$ $&$ $&$ $&&$$&$$&$$\\ 
$even\,II \; {}^{II}{\cal A}^{m}_f$&$1$& $[-i](+)(+) $ & $(-i)[+](+) $ & $[-i][+][+] $ & 
$(-i)(+)[+] $ &$-\frac{i}{2}$&
$\frac{1}{2}$&$\frac{1}{2}$\\ 
$$&$2$&    $(+i)[-](+) $ & $[+i](-)(+) $ & $(+i)(-)[+] $ & $[+i][-][+] $ &$\frac{i}{2}$&
$-\frac{1}{2}$&$\frac{1}{2}$ \\ 
$$&$3$&    $(+i)(+)[-] $ & $[+i][+][-] $ & $(+i)[+](-) $ & $[+i](+)(-) $&$\frac{i}{2}$&
$\frac{1}{2}$&$-\frac{1}{2}$\\ 
$$&$4$&    $[-i][-][-] $ & $(-i)(-)[-] $ & $[-i](-)(-) $ & $(-i)[-](-) $&$-\frac{i}{2}$&
$-\frac{1}{2}$&$-\frac{1}{2}$\\ 
\hline
 \end{tabular}
\end{center}
\end{tiny}
\end{table*}

While the Clifford odd ''basis vectors'' are (chosen to be) right handed, 
$\Gamma^{(5+1)}= 1$, have their Hermitian conjugated partners opposite 
handedness~\footnote{The handedness $\Gamma^{(d)}$, one of the invariants of 
the group $SO(d)$, 
with the infinitesimal generators of the Lorentz group $S^{ab}$, is
defined as 
$\Gamma^{(d)}=\alpha\, \varepsilon_{a_1 a_2\dots a_{d-1} a_d}\, S^{a_1 a_2} 
\cdot S^{a_3 a_4} \cdots S^{a_{d-1} a_d}$, with $\alpha$ chosen so that 
$\Gamma^{(d)}=\pm 1$.}
%


While the Clifford odd ''basis vectors'' have half integer eigenvalues of the Cartan subalgebra
members, Eq.(\ref{cartangrasscliff}), that is of $S^{03}, S^{12}, S^{56}$ in this particular 
case of $d=(5+1)$,  the Clifford even ''basis vectors'' have integer spins, obtained by 
${\bf {\cal S}}^{03}= S^{03}+ \tilde{S}^{03}$, ${\bf {\cal S}}^{12}= 
S^{12} +\tilde{S}^{12}$, 
${\bf {\cal S}}^{56}= S^{56}+ \tilde{S}^{56}$.

Let us check what does the algebraic application, $*_A$, of 
${}^{I}{\hat{\cal A}}^{m \dagger}_{f=3}, m=(1,2,3,4)$, presented in 
Table~\ref{Table Clifffourplet.} in the third column of $even\,I$, do on the Clifford odd 
''basis vectors'' $\hat{b}^{m=1 \dagger}_{f=1}$, presented as the first $odd \,I$ ''basis
vector'' in Table~\ref{Table Clifffourplet.}. This can easily be evaluated by taking into 
account Eq.~(\ref{graficcliff}) for any $m$.

\begin{small}
\begin{eqnarray}
&&{}^{I}{\hat{\cal A}}^{m \dagger}_{3}*_A \hat{b}^{1 \dagger}_{1} 
(\equiv \stackrel{03}{(+i)} \stackrel{12}{[+]} \stackrel{56}{[+]}):\nonumber\\
&&{}^{I}{\hat{\cal A}}^{1 \dagger}_{3} (\equiv \stackrel{03}{[+i]}
\stackrel{12}{[+]} \stackrel{56}{[+]})  *_{A} \hat{b}^{1 \dagger}_{1} 
(\equiv \stackrel{03}{(+i)} \stackrel{12}{[+]} \stackrel{56}{[+]}) \rightarrow
\hat{b}^{1 \dagger}_{1}\,,
{\rm   selfadjoint}\nonumber\\
&&{}^{I}{\hat{\cal A}}^{2 \dagger}_{3} (\equiv \stackrel{03}{(-i)}
\stackrel{12}{(-)} \stackrel{56}{[+]}) *_{A} \hat{b}^{1 \dagger}_{1}
\rightarrow \hat{b}^{2 \dagger}_{1} 
(\equiv \stackrel{03}{[-i]} \stackrel{12}{(-)} \stackrel{56}{[+]})\,, \quad  
\nonumber\\ 
&& {}^{I}{\hat{\cal A}}^{3 \dagger}_{3} (\equiv \stackrel{03}{(-i)}
\stackrel{12}{[+]} \stackrel{56}{(-)}) *_{A} \hat{b}^{1 \dagger}_{1}
\rightarrow \hat{b}^{3 \dagger}_{1} 
(\equiv \stackrel{03}{[-i]} \stackrel{12}{[+]} \stackrel{56}{(-)})\,, \quad  
\nonumber\\
&&{}^{I}{\hat{\cal A}}^{4 \dagger}_{3} (\equiv \stackrel{03}{[+i]}
\stackrel{12}{(-)} \stackrel{56}{(-)}) *_{A} \hat{b}^{1 \dagger}_{1}
\rightarrow \hat{b}^{4 \dagger}_{1}
(\equiv \stackrel{03}{(+i)} \stackrel{12}{(-)} \stackrel{56}{(-)})\,.\;  
\label{calAb1}
\end{eqnarray}
\end{small}
%
The sign $\rightarrow$ means that the relation is valid up to the constant.
${}^{I}{\hat{\cal A}}^{1 \dagger}_{3}$ is self adjoint, the Hermitian conjugated
partner of ${}^{I}{\hat{\cal A}}^{2 \dagger}_{3}$ is 
${}^{I}{\hat{\cal A}}^{1 \dagger}_{4}$, of ${}^{I}{\hat{\cal A}}^{3 \dagger}_{3}$
is ${}^{I}{\hat{\cal A}}^{1 \dagger}_{2}$ and of 
${}^{I}{\hat{\cal A}}^{4 \dagger}_{3}$ is ${}^{I}{\hat{\cal A}}^{1 \dagger}_{1}$.

We can conclude that the  algebraic, $*_A$,  application  of
 ${}^{I}{\hat{\cal A}}^{m \dagger}_{3} (\equiv \stackrel{03}{(-i)}
\stackrel{12}{[+]} \stackrel{56}{(-)}) $ on $\hat{b}^{1 \dagger}_{1}$ leads to
the same or another family member of the same family $f=1$, namely to
$\hat{b}^{m \dagger}_{1} $, $m=(1,2,3,4)$.

Calculating the eigenvalues  of the Cartan subalgebra members, Eq.~(\ref{cartangrasscliff}), before and after the algebraic multiplication, $*_A$, one sees that 
${}^{I}{\hat{\cal A}}^{m \dagger}_{3}$ carry the integer eigenvalues of the Cartan 
subalgebra members, namely of ${\cal {\bf S}}^{ab}$ $= S^{ab} + \tilde{S}^{ab} $,
since they transfer when applying on the  Clifford odd ''basis vector'' to it the 
integer eigenvalues  of the Cartan subalgebra members, changing  the  Clifford odd 
''basis vector'' into another Clifford odd ''basis vector''.

We  therefore find out that the algebraic application 
 of ${}^{I}{\hat{\cal A}}^{m \dagger}_{3} $, $m=1,2,3,4$,
on $\hat{b}^{1 \dagger}_{1}$ transforms $\hat{b}^{1 \dagger}_{1}$ into
$\hat{b}^{m \dagger}_{1}$, $m=(1,2,3,4)$.
Similarly we find that the algebraic application of ${}^{I}{\hat {\cal A}}^{m}_4,$ 
$m=(1,2,3,4)$ on $\hat{b}^{2 \dagger}_{1}$ transforms $\hat{b}^{2 \dagger}_{1}$ 
into $\hat{b}^{m \dagger}_{1}, m=(1,2,3,4)$.
The algebraic application of ${}^{I}{\hat {\cal A}}^{m}_2,$ $m=(1,2,3,4)$ on   
$\hat{b}^{3 \dagger}_{1}$ transforms $\hat{b}^{3 \dagger}_{1}$ into
$\hat{b}^{m \dagger}_{1}, m=(1,2,3,4)$.
And the algebraic application of ${}^{I}{\hat {\cal A}}^{m}_1,$ $m=(1,2,3,4)$ on   
$\hat{b}^{4 \dagger}_{1}$ transforms $\hat{b}^{4 \dagger}_{1}$ into
$\hat{b}^{m \dagger}_{1}, m=(1,2,3,4)$.

\vspace{2mm}
 
 The statement of Eq.~(\ref{calAb1234gen}) is therefore demonstrated on the case
 of $d=(5+1)$.

It remains to stress  and illustrate in the case of $d=(5+1)$ some general properties of 
the Clifford even ''basis vector'' 
${}^{I}{\hat{\cal A}}^{m \dagger}_{f}$ when they 
apply on each other. Let us denote the self adjoint member in each group of
''basis vectors'' of particular $f$  as ${}^{I}{\hat{\cal A}}^{m_{0} \dagger}_{f}$.
 We easily see that
\begin{eqnarray}
\label{evenproperties}
\{{}^{I}{\hat{\cal A}}^{m \dagger}_{f}\,, 
{}^{I}{\hat{\cal A}}^{m' \dagger}_{f}\}_{-}&=&0\,,
\quad {\rm if }\,  (m,m') \, \ne m_0  \,{\rm or}\, m=m_0=m'\,, \forall \, f\,,\nonumber\\
{}^{I}{\hat{\cal A}}^{m \dagger}_{f}*_{A} {}^{I}{\hat{\cal A}}^{m_0 \dagger}_{f}
&\rightarrow&
{}^{I}{\hat{\cal A}}^{m \dagger}_{f}\,, \quad \forall \, m \,, \,\forall \,f\,.
\end{eqnarray}

Since a projector to the second power is the projector  back, and since a projector is 
self adjoint,  it means that the self adjoint Clifford even ''basis vectors'' is the  product of projectors. 
 In Table~\ref{Table Clifffourplet.} we see that in each column of either 
 $even \,{}^{I}{\hat{\cal A}}^{m \dagger}_{f}$ or of  
 $even {}^{II}{\hat{\cal A}}^{m \dagger}_{f}$ there is one self adjoint 
 ${}^{I,II}{\hat{\cal A}}^{m_{0}\dagger}_{f}$. We also see that
two ''basis vectors'' ${}^{I}{\hat{\cal A}}^{m \dagger}_{f}$  and 
${}^{I}{\hat{\cal A}}^{m ' \dagger}_{f}$ of the same $f$ and of $(m, m')\ne m_0$ are orthogonal. We only have to take into account Eq.~(\ref{graficcliff1}), which tells that
$$\stackrel{ab}{(k)}\stackrel{ab}{[k]}= 0\,,\quad 
\stackrel{ab}{[k]}\stackrel{ab}{(k)}=  \stackrel{ab}{(k)}\,, \quad 
  \stackrel{ab}{(k)}\stackrel{ab}{[-k]} =  \stackrel{ab}{(k)}\,,
\quad \, \stackrel{ab}{[k]}\stackrel{ab}{(-k)} =0.$$

\noindent
These relations tell us that 
${}^{I}{\hat{\cal A}}^{1 \dagger}_{4}$ $ *_{A} {}^{I}{\hat{\cal A}}^{2 \dagger}_{3} $
$= {}^{I}{\hat{\cal A}}^{1 \dagger}_{3}$,
what illustrates Eq.~(\ref{ruleAAI3}), while 
${}^{I}{\hat{\cal A}}^{2 \dagger}_{3}$$ *_{A} {}^{I}{\hat{\cal A}}^{1 \dagger}_{4} $
$= {}^{I}{\hat{\cal A}}^{2 \dagger}_{4}$ 
illustrating  Eq.~(\ref{ruleAAI2}), while 
${}^{I}{\hat{\cal A}}^{1 \dagger}_{3}$
$ *_{A} {}^{I}{\hat{\cal A}}^{2 \dagger}_{4} =0$ illustrates  Eq.~(\ref{ruleAAI1}).\\

Table~\ref{Cliff basis5+1even I.} presents the Clifford even 
''basis vectors'' ${}^{I}{\hat{\cal A}}^{m \dagger}_{f}$ for $d=(5+1)$ with the properties:
$\;\;\;$ {\bf i.}They are products of an even number of nilpotents,  $\stackrel{ab}{(k)}$, 
with the rest up to $\frac{d}{2}$ of projectors,~$\stackrel{ab}{[k]}$.
$\;\;\;$ {\bf ii.}~Nilpotents and projectors are eigenvectors of the Cartan 
subalgebra members ${\bf {\cal S}}^{ab}$ $= S^{ab} + \tilde{S}^{ab} $, Eq.~(\ref{cartangrasscliff}), carrying the integer eigenvalues 
of the Cartan subalgebra members.\\
$\;\;\;$ {\bf iii.} They have their Hetmitian conjugated partners within the same group  of
${}^{I}{\hat{\cal A}}^{m \dagger}_{f}$ with $2^{\frac{d}{2}-1}$ 
$\times$ $2^{\frac{d}{2}-1}$ members. \\
$\;\;\;$ {\bf iv.} They have properties of the boson gauge fields. 
When applying on the Clifford odd ''basis vectors'' (offering the description of 
the fermion fields) they transform the Clifford odd ''basis vectors'' into
another Clifford odd ''basis vectors'', transferring to the Clifford odd ''basis vectors'' 
the integer spins with respect to the  $SO(d-1,1)$ group, 
while with respect to subgroups of the  $SO(d-1,1)$ group they transfer appropriate
superposition of the eigenvalues (manifesting the properties of the adjoint 
representations of the corresponding groups).\\

To  demonstrate that the  Clifford even ''basis vectors'' have properties of the gauge 
fields of the corresponding  Clifford odd ''basis vectors'' we study properties of the
 $SU(3)$ $\times U(1)$ subgroups of the Clifford odd and Clifford even ''basis vectors''. 

\begin{small}
We present in Eqs.~(\ref{so1+3 5+1}, \ref{so64 5+1}) the superposition of members 
of Cartan subalgebra, Eq.~(\ref{cartangrasscliff}) for $S^{ab}$ for the Clifford odd
''basis  vectors'',  for the subgroup  $SO(3,1) \times U(1)$ ($N^3_{\pm}\,,\tau$)  and 
for the subgroup  $SU(3)$ $\times U(1)$: ($\tau',\tau^{3}, \tau^{8}$).  
The same relations can be used also for the corresponding operators determining the 
''family'' quantum numbers ($\tilde{N}^3_{\pm}\,, \tilde{\tau}$) of the Clifford odd
''basis  vectors', if  $S^{ab}$'s are  replaced by $\tilde{S}^{ab}$'s. For the Clifford 
even objects ${\cal {\bf S}}^{ab}(=S^{ab}+ \tilde{S}^{ab})$ must replace $S^{ab}$.
\begin{eqnarray}
\label{so1+3 5+1}
&& N^3_{\pm}(= N^3_{(L,R)}): = \,\frac{1}{2} (%
 S^{12}\pm i S^{03} )\,,\quad \tau =S^{56}\,, \quad
\end{eqnarray}
%
%
 \begin{eqnarray}
 \label{so64 5+1}
 \tau^{3}: = &&\frac{1}{2} \,(%
 -S^{1\,2} - iS^{0\,3})\, , \qquad 
\tau^{8}= \frac{1}{2\sqrt{3}} (-i S^{0\,3} + S^{1\,2} -  2 S^{5\;6})\,,\nonumber\\
 \tau' = &&-\frac{1}{3}(-i S^{0\,3} + S^{1\,2} + S^{5\,6})\,.
%
 \end{eqnarray}
\end{small}

\begin{figure}
  \centering
   \includegraphics[width=0.45\textwidth]{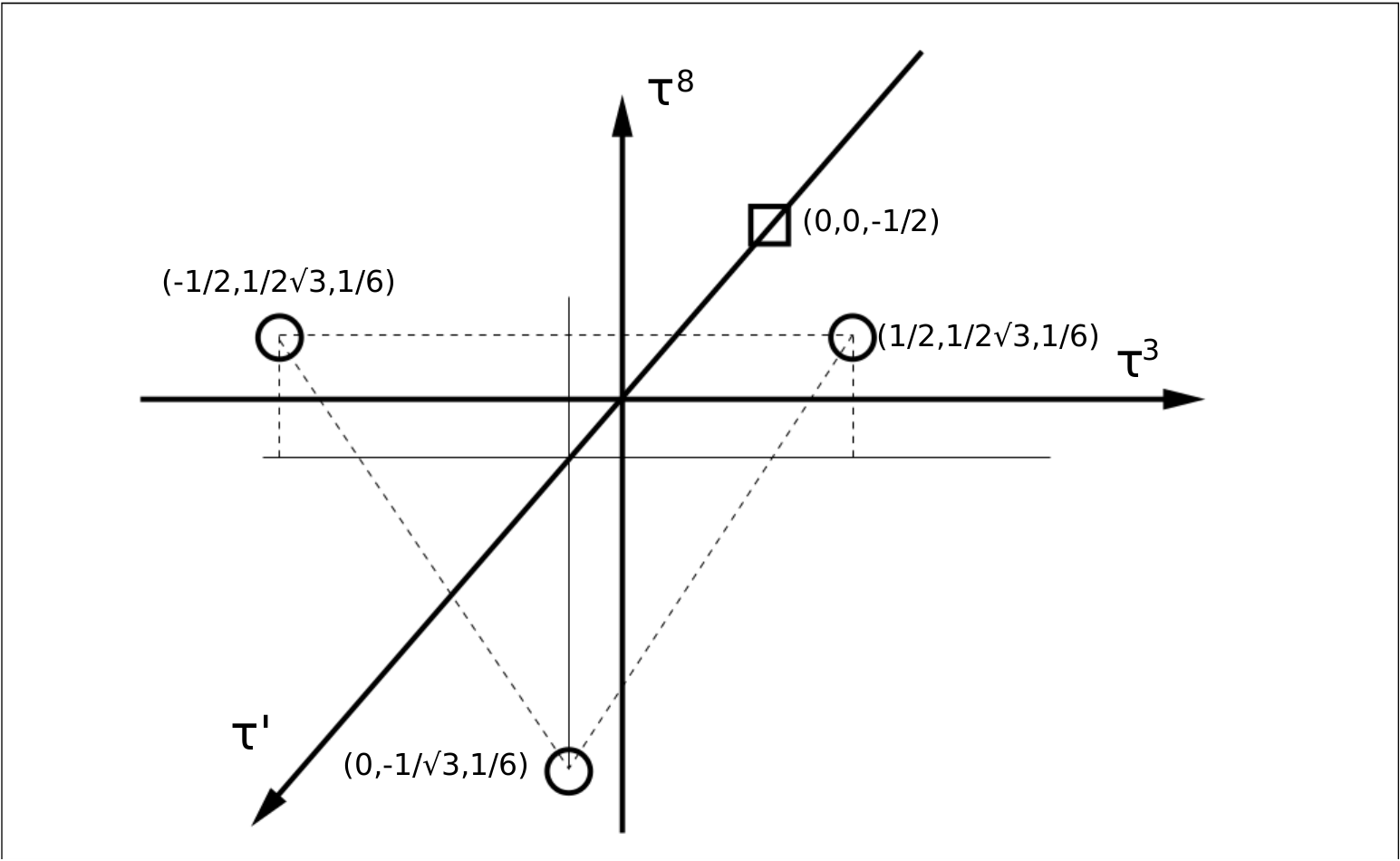}
  \caption{\label{FigSU3U1odd} 
  Representations of the subgroups $SU(3)$ and $U(1)$ of the group 
$SO(5,1)$, the properties of which appear in  Table~\ref{Table Clifffourplet.}, are 
presented. ($\tau^3, \tau^8$ and $\tau'$)  can be calculated if using 
Eqs.(\ref{so1+3 5+1}, \ref{so64 5+1}).
On the abscissa axis, on the ordinate axis and on the third axis the eigenvalues of the  superposition of the three Cartan subalgebra members, $\tau^3$%
, $\tau^8$
, $\tau'$ 
are presented. One notices one triplet,
denoted by ${\bf \bigcirc}$ with the values $\tau'=\frac{1}{6}$, ($\tau^3=-\frac{1}{2},
\tau^8=\frac{1}{2\sqrt{3}}, \tau'=\frac{1}{6})$, ($\tau^3=\frac{1}{2},
\tau^8=\frac{1}{2\sqrt{3}}, \tau'=\frac{1}{6}$), ($\tau^3=0,
\tau^8=-\frac{1}{\sqrt{3}}, \tau'=\frac{1}{6}$), respectively, and one singlet denoted 
by the square. 
($\tau^3=0, \tau^8=0, \tau'=-\frac{1}{2}$). 
The triplet and the singlet appear  in four families.} 
\end{figure}
\begin{figure}
  \centering
   \includegraphics[width=0.45\textwidth]{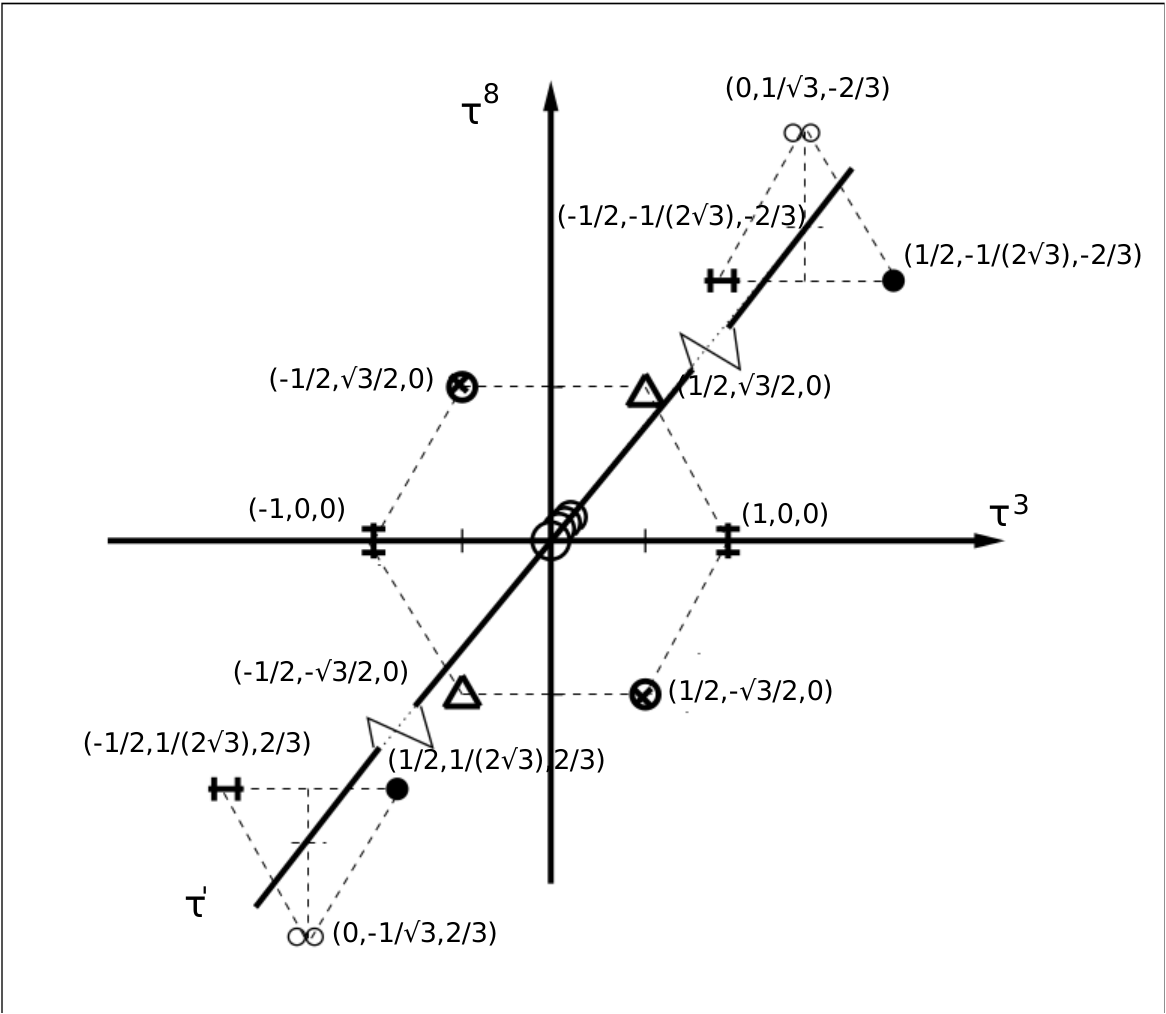}
  \caption{\label{FigSU3U1even} 
The Clifford even ''basis vectors'' ${}^{I}{\hat{\cal A}}^{m}_{f}$, in the case that 
$d=(5+1)$, are presented with respect to the eigenvalues of the commuting operators 
of the subgroups $SU(3)$ and $U(1)$ of the group $SO(5,1)$: $\tau^3=\frac{1}{2}$
$(- {\cal S}^{12} -i {\cal S}^{03})$, $\tau^8=\frac{1}{2\sqrt{3}} 
({\cal S}^{12} -i {\cal S}^{03}- 2 {\cal S}^{56})$, 
$\tau'=- \frac{1}{3} ({\cal S}^{12} -i {\cal S}^{03} + {\cal S}^{56})$. 
Their properties appear  in  Table~\ref{Cliff basis5+1even I.}. 
The abscissa axis carries the eigenvalues of $\tau^3$, the ordinate axis of $\tau^8$
and the third axis the eigenvalues of $\tau'$,
One notices four singlets with ($\tau^3=0, \tau^8=0, \tau'=0$), denoted by $\bigcirc$, representing four self adjoint  Clifford even ''basis vectors'' ${}^{I}{\hat{\cal A}}^{m}_{f}$, 
one sextet of three pairs with $\tau'=0$, Hermitian conjugated to each other, denoted by 
$\bigtriangleup$ (with  ($\tau'=0, \tau^3=-\frac{1}{2}, \tau^8=-\frac{3}{2\sqrt{3}}$) 
and ($\tau'=0, \tau^3=\frac{1}{2}, \tau^8=\frac{3}{2\sqrt{3}}$) ), respectively, 
by $\ddagger$ (with  ($\tau'=0, \tau^3=-1, \tau^8=0$) 
and ($\tau'=0, \tau^3=1, \tau^8=0$), respectively, and by $\otimes$ 
(with  ($\tau'=0, \tau^3=\frac{1}{2}, \tau^8=-\frac{3}{2\sqrt{3}}$) 
and ($\tau'=0, \tau^3=- \frac{1}{2}, \tau^8=\frac{3}{2\sqrt{3}}$) ), respectively, 
and one triplet, denoted by $\star \star$ with ($\tau'=\frac{2}{3}, \tau^3=\frac{1}{2}, \tau^8=\frac{1}{2\sqrt{3}}$), by $\bullet$ with  ($\tau'=\frac{2}{3}, 
\tau^3= -\frac{1}{2}, \tau^8=\frac{1}{2\sqrt{3}}$), and by $\odot \odot$ with
 ($\tau'=\frac{2}{3}, \tau^3=0, \tau^8=-\frac{1}{\sqrt{3}}$), as well as one
 antitriplet, Hermitian conjugated to the triplet, denoted by $\star \star$ with 
 ($\tau'=-\frac{2}{3}, \tau^3=-\frac{1}{2}, \tau^8=-\frac{1}{2\sqrt{3}}$), 
 by $\bullet$ with  ($\tau'=-\frac{2}{3}, \tau^3= \frac{1}{2}, \tau^8=- 
 \frac{1}{2\sqrt{3}}$), and by $\odot \odot$ with  ($\tau'=-\frac{2}{3}, 
 \tau^3=0, \tau^8=\frac{1}{\sqrt{3}}$).} 
\end{figure}

Let us, for example, algebraically apply ${}^{I}{\hat{\cal A}}^{2}_{3}$ 
($\equiv \stackrel{03}{(-i)}\,\stackrel{12}{(-)}\,\stackrel{56}{[+]}$), denoted 
by $\odot\odot$ on Table~\ref{Cliff basis5+1even I.}, carrying $(\tau^3=0, 
\tau^8= - \frac{1}{\sqrt 3}, \tau'=\frac{2}{3})$, represented also on 
Fig.~\ref{FigSU3U1even} by $\odot\odot$,  on the Clifford odd ''basis vector''  
$\hat{b}^{1\dagger}_{1} (\equiv \stackrel{03}{(+i)} \stackrel{12}{(+)} 
\stackrel{56}{(+)}) $, presented on Table~\ref{Table Clifffourplet.}, with  
$(\tau^3=0, \tau^8= 0, \tau'=-\frac{1}{2})$, as we can calculate using 
Eq.~(\ref{so64 5+1}) and which is represented on Fig.~\ref{FigSU3U1odd} 
by a square  as a singlet. ${}^{I}{\hat{\cal A}}^{2}_{3}$ 
transforms $\hat{b}^{1\dagger}_{1}$  (by transferring to $\hat{b}^{1\dagger}_{1}$ 
$(\tau^3=0, \tau^8= - \frac{1}{\sqrt 3}, \tau'=\frac{2}{3})$)  to $\hat{b}^{1\dagger}_{2}$ with $(\tau^3=0, \tau^8= - \frac{1}{\sqrt{3}}, \tau'=\frac{1}{6})$, belonging 
on Fig.~\ref{FigSU3U1odd} to the triplet, denoted by $\bigcirc$. The corresponding gauge 
fields, presented on Fig.~\ref{FigSU3U1even}, if belonging to the sextet, would transform
the triplet of quarks among themselves. 

%
In the case of the group $SO(6)$, manifesting as $SU(3) \times U(1)$ and representing the
$SU(3)$ colour group and $U(1)$ the ''fermion'' quantum number, embedded into 
$SO(13,1)$ the triplet would represent quarks and the singlet leptons.
The corresponding gauge of the fields, presented on Fig.~\ref{FigSU3U1even}, if 
belonging to the sextet, would transform the triplet of quarks among themselves, changing
the colour and leaving the ''fermion'' quantum number equal to $\frac{1}{6}$. \\

%

We can see that ${}^{I}{\hat{\cal A}}^{m \dagger}_{3}$  with $(m=2,3,4)$,
if applied on the $SU(3)$ singlet $\hat{b}^{1 \dagger}_{1}$ with ($\tau'= -\frac{1}{2}, \tau^3=0,\tau^8=0$), transforms it to  $\hat{b}^{m=2,3,4) \dagger}_{1}$, respectively, which are members of the $SU(3 )$ triplet. All these Clifford even 
''basis vectors'' have $\tau'$ equal to $\frac{2}{3}$, changing correspondingly 
$\tau'= -\frac{1}{2}$ into 
$\tau'=\frac{1}{6}$ and  bringing the needed values of $\tau^3$ and $\tau^8$. 

In Table~\ref{Cliff basis5+1even I.} we find $(6+4)$ Clifford even ''basis vectors'' 
${}^{I}{\hat{\cal A}}^{m \dagger}_{f}$ with $\tau `=0$. 
Six of them are Hermitian 
conjugated to each other --- the Hermitian conjugated partners are denoted by the 
same geometric figure on the third column. Four of them are self adjoint and 
correspondingly with ($\tau '=0, \tau^3=0, \tau^8=0$), denoted in the third 
column of Table~\ref{Cliff basis5+1even I.} by $\bigcirc$.
The rest $6$  Clifford even ''basis vectors'' belong to one triplet with $\tau '=\frac{2}{3}$ 
and $(\tau^3, \tau^8)$ equal to [$(0,- \frac{1}{\sqrt{3}})$, $(-\frac{1}{2}, 
\frac{1}{2\sqrt{3}})$, $(\frac{1}{2}, \frac{1}{2\sqrt{3}})$] and one antitriplet 
with $\tau '=-\frac{2}{3}$ and ($(\tau^3, \tau^8)$ equal to [$(-\frac{1}{2}, 
-\frac{1}{2\sqrt{3}})$, $(\frac{1}{2}, -\frac{1}{2\sqrt{3}})$, 
$(0, \frac{1}{\sqrt{3}})$]. 

Each triplet has Hermitian conjugated partners in antitriplet and opposite. In 
Table~\ref{Cliff basis5+1even I.} the Hermitian conjugated partners of the triplet and 
antitriplet are denoted by the same signum:  (${}^{I}{\hat{\cal A}}^{1 \dagger}_{1}$, 
${}^{I}{\hat{\cal A}}^{4 \dagger}_{3}$) by
$\star \star$, (${}^{I}{\hat{\cal A}}^{1 \dagger}_{2}$, 
 ${}^{I}{\hat{\cal A}}^{3 \dagger}_{3}$)  by $\bullet$, and 
 (${}^{I}{\hat{\cal A}}^{2 \dagger}_{3}$, ${}^{I}{\hat{\cal A}}^{1 \dagger}_{4}$) 
by $\odot \odot$.

The octet and the two triplets are presented in Fig.~\ref{FigSU3U1even}.

\vspace{2mm}

%
\begin{table}
\begin{tiny}
\caption{The  Clifford even ''basis vectors''  ${}^{I}{\hat{\cal A}}^{m \dagger}_{f}$,  
each of them is the product of projectors and an even number of nilpotents, and each is 
the eigenvector of all the Cartan subalgebra members, ${\cal S}^{03}$, 
${\cal S}^{12}$, ${\cal S}^{56}$, Eq.~(\ref{cartangrasscliff}), are presented for 
$d= (5+1)$-dimensional case. Indexes $m$ and $f$ determine  
$2^{\frac{d}{2}-1}\times 2^{\frac{d}{2}-1}$ different members  
${}^{I}{\hat{\cal A}}^{m \dagger}_{f}$. 
In the third column  the  ''basis vectors'' ${}^{I}{\hat{\cal A}}^{m \dagger}_{f}$  
which are  Hermitian conjugated partners to each other (and can therefore annihilate 
each other) are pointed out with the same symbol. For example, with $\star \star$ 
are equipped the first member with $m=1$ and $f=1$ and the last member of $f=3$ 
with $m=4$.
The sign $\bigcirc$ denotes the  Clifford even ''basis vectors'' which are self adjoint 
$({}^{I}{\hat{\cal A}}^{m \dagger}_{f})^{\dagger}$ 
$={}^{I}{\hat{\cal A}}^{m' \dagger}_{f `}$. It is obvious that ${}^{\dagger}$
has no meaning, since ${}^{I}{\hat{\cal A}}^{m \dagger}_{f}$ are self adjoint or 
are Hermitian conjugated partner to another ${}^{I}{\hat{\cal A}}^{m' \dagger}_{f `}$.
This table represents also the  eigenvalues of the three commuting operators 
${\cal N}^3_{L,R}$ and ${\cal S}^{56 }$ of the subgroups
 $SU(2)\times SU(2)\times U(1)$ of the group
$SO(5,1)$ and the eigenvalues of the three 
commuting operators ${\tau}^3, {\tau}^8$ and ${ \tau'}$ of the 
subgroups  $SU(3)\times U(1)$.
\vspace{3mm}
}
\label{Cliff basis5+1even I.} 
 %
 \begin{center}
 \begin{tabular}{|r| r|r|r|r|r|r|r|r|r|r|r|}
 \hline
$\, f $&$m $&$*$&${}^{I}\hat{\cal A}^{m \dagger}_f$
&${\cal S}^{03}$&$ {\cal S}^{1 2}$&${\cal S} ^{5 6}$&
${\cal N}^3_L$&${\cal N}^3_R$&
${\cal \tau}^3$&${\cal \tau}^8$&${\cal \tau}'$
\\
\hline
%
$I$&$1$&$\star \star$&$
\stackrel{03}{[+i]}\,\stackrel{12}{(+)} \stackrel{56}{(+)}$&
$0$&$1$&$1$.
&$\frac{1}{2}$&$\frac{1}{2}$&$-\frac{1}{2}$&$-\frac{1}{2\sqrt{3}}$&$-\frac{2}{3}$
\\
$$ &$2$&$\bigtriangleup$&$
\stackrel{03}{(-i)}\,\stackrel{12}{[-]}\,\stackrel{56}{(+)}$&
$- i$&$0$&$1$
&$\frac{1}{2}$&$-\frac{1}{2}$&$-\frac{1}{2}$&$-\frac{3}{2\sqrt{3}}$&$0$
\\
$$ &$3$&$\ddagger$&$
\stackrel{03}{(-i)}\,\stackrel{12}{(+)}\,\stackrel{56}{[-]}$&
$-i$&$ 1$&$0$
&$1 $&$0$&$-1$&$0$&$0$
\\
$$ &$4$&$\bigcirc$&$
\stackrel{03}{[+i]}\,\stackrel{12}{[-]}\,\stackrel{56}{[-]}$&
$0$&$0$&$0$
&$0$&$0$&$0$&$0$&$0$
\\
\hline 
$II$&$1$&$\bullet$&$
\stackrel{03}{(+i)}\,\stackrel{12}{[+]}\, \stackrel{56}{(+)}$&
$i$&$0$&$1$&
$-\frac{1}{2}$&$\frac{1}{2}$&$\frac{1}{2}$&$-\frac{1}{2\sqrt{3}}$&$-\frac{2}{3}$\\
$$ &$2$&$\otimes$&$
\stackrel{03}{[-i]}\,\stackrel{12}{(-)}\,\stackrel{56}{(+)}$&
$0$&$-1$&$1$
&$-\frac{1}{2}$&$-\frac{1}{2}$&$\frac{1}{2}$&$-\frac{3}{2\sqrt{3}}$&$0$
\\
$$ &$3$&$\bigcirc$&$
\stackrel{03}{[-i]}\,\stackrel{12}{[+]}\,\stackrel{56}{[-]}$&
$0$&$ 0$&$0$
&$0$&$0$&$0$&$0$&$0$
\\
$$ &$4$&$\ddagger$&$
\stackrel{03}{(+i)}\, \stackrel{12}{(-)}\,\stackrel{56}{[-]}$&
$i$&$-1$&$0$
&$-1$&$0$&$1$&$0$&$0$
\\ 
%
%
 \hline
$III$&$1$&$\bigcirc$&$
\stackrel{03}{[+i]}\,\stackrel{12}{[+]}\, \stackrel{56}{[+]}$&
$0$&$0$&$0$&
$0$&$0$&$0$&$0$&$0$\\
$$ &$2$&$\odot \odot$&$
\stackrel{03}{(-i)}\,\stackrel{12}{(-)}\,\stackrel{56}{[+]}$&
$-i$&$-1$&$0$
&$0$&$-1$&$0$&$-\frac{1}{\sqrt{3}}$&$\frac{2}{3}$\\
$$ &$3$&$\bullet$&$
\stackrel{03}{(-i)}\,\stackrel{12}{[+]}\,\stackrel{56}{(-)}$&
$-i$&$ 0$&$-1$
&$\frac{1}{2}$&$-\frac{1}{2}$&$-\frac{1}{2}$&$\frac{1}{2\sqrt{3}}$&$\frac{2}{3}$
\\
$$ &$4$&$\star \star$&$
\stackrel{03}{[+i]} \stackrel{12}{(-)}\,\stackrel{56}{(-)}$&
$0$&$- 1$&$- 1$
&$-\frac{1}{2}$&$-\frac{1}{2}$&$\frac{1}{2}$&$\frac{1}{2\sqrt{3}}$&$\frac{2}{3}$
\\
\hline
$IV$&$1$&$\odot \odot $&$
\stackrel{03}{(+i)}\,\stackrel{12}{(+)}\, \stackrel{56}{[+]}$&
$i$&$1$&$0$&
$0$&$1$&$0$&$\frac{1}{\sqrt{3}}$&$-\frac{2}{3}$
\\
$$ &$2$&$\bigcirc$&$
\stackrel{03}{[-i]}\,\stackrel{12}{[-]}\,\stackrel{56}{[+]}$&
$0$&$0$&$0$
&$0$&$0$&$0$&$0$&$0$
\\
$$ &$3$&$\otimes$&$
\stackrel{03}{[-i]}\,\stackrel{12}{(+)}\,\stackrel{56}{(-)}$&
$0$&$ 1$&$-1$
&$\frac{1}{2}$&$\frac{1}{2}$&$-\frac{1}{2}$&$\frac{3}{2\sqrt{3}}$&$0$
\\
$$ &$4$&$\bigtriangleup$&$
\stackrel{03}{(+i)}\, \stackrel{12}{[-]}\,\stackrel{56}{(-)}$&
$i$&$0$&$-1$
&$-\frac{1}{2}$&$\frac{1}{2}$&$\frac{1}{2}$&$\frac{3}{2\sqrt{3}}$&$0$\\ 
\hline 
 \end{tabular}
 \end{center}
\end{tiny}
\end{table}

Fig.~\ref{FigSU3U1even} represents the $2^{\frac{d}{2}-1}\times 2^{\frac{d}{2}-1}$ 
members  ${}^{I}{\hat{\cal A}}^{m}_{f}$ of the Clifford even ''basis vectors'' for the
case that $d=(5+1)$. The properties of ${}^{I}{\hat{\cal A}}^{m}_{f}$ are presented 
also in Table~\ref{Cliff basis5+1even I.}. There are in this case  again $16$ members. Manifesting the structure of subgroups $SU(3) \times U(1)$ of the group $SO(5,1)$ they 
are represented as eigenvectors of the superposition of the Cartan subalgebra members 
(${\bf {\cal S}}^{03}, {\bf {\cal S}}^{12}, {\bf {\cal S}}^{56}$), that is with 
$\tau^3=\frac{1}{2} (- {\bf {\cal S}}^{12} -i {\bf {\cal S}}^{03})$, 
$\tau^8=\frac{1}{2\sqrt{3}} ( {\bf {\cal S}}^{12} -i {\bf {\cal S}}^{03}- 
2 {\bf {\cal S}}^{56})$, and $\tau'=- \frac{1}{3} 
({\bf {\cal S}}^{12} -i {\bf {\cal S}}^{03} + {\bf {\cal S}}^{56})$. 
There are four self adjoint Clifford even ''basis vectors'' with ($\tau^3=0, \tau^8=0, \tau'=0$),
one sextet of three pairs Hermitian conjugated to each other, one triplet and one 
antitriplet with the members of the triplet Hermitian conjugated to the corresponding 
members of the antitriplet and opposite. These $16$ members of the Clifford even 
''basis vectors'' ${}^{I}{\hat{\cal A}}^{m}_{f}$ are the boson ''partners'' of the 
Clifford odd ''basis vectors'' $\hat{b}^{m \dagger }_{f}$, presented in Fig.~\ref{FigSU3U1odd} for one of four families, anyone.  The reader can check 
that the algebraic application of ${}^{I}{\hat{\cal A}}^{m}_{f}$,  belonging to 
the triplet, transforms the Clifford odd singlet, denoted on Fig.~\ref{FigSU3U1odd} 
by a square, to one of the members of the triplet, denoted  on Fig.~\ref{FigSU3U1odd} 
by  the circle ${\bf \bigcirc}$.

Looking at the boson fields ${}^{I}{\hat{\cal A}}^{m \dagger}_{f}$ from the point of view
of subgroups  $SU(3)\times U(1)$ of the group $SO(5+1)$ we will recognize in the part 
of fields forming the octet the colour gauge fields of quarks and leptons and antiquarks and antileptons.

\vspace{1mm}
\subsection{Second quantized fermion and boson fields the internal spaces of which are
described by the Clifford basis vectors.}
\label{secondquantizedfermionsbosons}

\vspace{2mm}

We learned in the previous sub section that in  even dimensional spaces ($d=2(2n+1)$ 
or $d=4n$) the Clifford odd and the Clifford even ''basis vectors'', which are the 
superposition of the Clifford odd and the Clifford even products of $\gamma^a$'s, 
respectively, offer the description of the internal spaces of fermion and boson fields.

The Clifford odd algebra offers $2^{\frac{d}{2}-1}$ ''basis vectors''  
$\hat{b}^{m \dagger}_{f}$, appearing in $2^{\frac{d}{2}-1}$ families 
(with the family quantum numbers determined by  $\tilde{S}^{ab}= \frac{i}{2} \{ \tilde{\gamma}^a, \tilde{\gamma}^b\}_{-}$),  which together with their 
$2^{\frac{d}{2}-1}\times$ $2^{\frac{d}{2}-1}$ Hermitian conjugated partners 
$\hat{b}^{m}_{f}$ fulfil the postulates for the second quantized fermion fields,  Eq.~(\ref{almostDirac}) in this paper, Eq.(26) in Ref.~\cite{nh2021RPPNP}, 
explaining the second quantization postulates of Dirac.

The Clifford even algebra offers $2^{\frac{d}{2}-1}\times$ $2^{\frac{d}{2}-1}$ 
''basis vectors'' of ${}^{I}{\hat{\cal A}}^{m \dagger}_{f}$ (and the same number
of ${}^{II}{\hat{\cal A}}^{m \dagger}_{f}$)  with the properties of the second 
quantized boson fields manifesting as the gauge fields of fermion fields described 
by the Clifford odd ''basis vectors'' $\hat{b}^{m \dagger}_{f}$.

The Clifford odd and the Clifford even ''basis vectors'' are chosen to be products of 
nilpotents, $\stackrel{ab}{(k)}$ (with the odd number of nilpotents if describing 
fermions and the even number of nilpotents if describing bosons), and projectors,  
$\stackrel{ab}{[k]}$. Nilpotents and projectors are (chosen to be) eigenvectors 
of the Cartan subalgebra members of the Lorentz algebra in the internal space of 
$S^{ab}$ for the Clifford odd ''basis vectors''  and of ${\bf {\cal S}}^{ab} (=
S^{ab}+ \tilde{S}^{ab}$) for  the Clifford even ''basis vectors''.

\vspace{3mm}

To define the creation operators, either for fermions or for bosons 
besides the ''basis vectors'' defining the internal space of fermions and bosons 
also the basis in ordinary space in momentum or coordinate representation is needed.
 Here  Ref.~\cite{nh2021RPPNP}, Subsect.~3.3 and App. J is overviewed. \\

Let us introduce the momentum  part of the  single particle states. The longer version 
is presented in Ref.~\cite{nh2021RPPNP} in Subsect.~3.3 and in App. J.
\begin{eqnarray}
\label{creatorp}
|\vec{p}>&=& \hat{b}^{\dagger}_{\vec{p}} \,|\,0_{p}\,>\,,\quad 
<\vec{p}\,| = <\,0_{p}\,|\,\hat{b}_{\vec{p}}\,, \nonumber\\
<\vec{p}\,|\,\vec{p}'>&=&\delta(\vec{p}-\vec{p}')=
<\,0_{p}\,|\hat{b}_{\vec{p}}\; \hat{b}^{\dagger}_{\vec{p}'} |\,0_{p}\,>\,, 
\nonumber\\
&&{\rm leading \;to\;} \nonumber\\
\hat{b}_{\vec{p'}}\, \hat{b}^{\dagger}_{\vec{p}} &=&\delta(\vec{p'}-\vec{p})\,,
\end{eqnarray}
with the normalization  $<\,0_{p}\, |\,0_{p}\,>=1$. 
While the quantized operators $\hat{\vec{p}}$ and  $\hat{\vec{x}}$ commute
 $\{\hat{p}^i\,, \hat{p}^j \}_{-}=0$ and  $\{\hat{x}^k\,, \hat{x}^l \}_{-}=0$, 
it follows for  $\{\hat{p}^i\,, \hat{x}^j \}_{-}=i \eta^{ij}$. One correspondingly 
finds 
\begin{small}
 \begin{eqnarray}
 \label{eigenvalue10}
 <\vec{p}\,| \,\vec{x}>&=&<0_{\vec{p}}\,|\,\hat{b}_{\vec{p}}\;
\hat{b}^{\dagger}_{\vec{x}} 
 |0_{\vec{x}}\,>=(<0_{\vec{x}}\,|\,\hat{b}_{\vec{x}}\;
\hat{b}^{\dagger}_{\vec{p}} \,
 |0_{\vec{p}}\,>)^{\dagger}\, \nonumber\\
 \{\hat{b}^{\dagger}_{\vec{p}}\,,  \,
\hat{b}^{\dagger}_{\vec{p}\,'}\}_{-}&=&0\,,\qquad 
\{\hat{b}_{\vec{p}},  \,\hat{b}_{\vec{p}\,'}\}_{-}=0\,,\qquad
\{\hat{b}_{\vec{p}},  \,\hat{b}^{\dagger}_{\vec{p}\,'}\}_{-}=0\,,
\nonumber\\
\{\hat{b}^{\dagger}_{\vec{x}},  \,\hat{b}^{\dagger}_{\vec{x}\,'}\}_{-}&=&0\,,
\qquad 
\{\hat{b}_{\vec{x}},  \,\hat{b}_{\vec{x}\,'}\}_{-}=0\,,\qquad
\{\hat{b}_{\vec{x}},  \,\hat{b}^{\dagger}_{\vec{x}\,'}\}_{-}=0\,,
\nonumber\\
\{\hat{b}_{\vec{p}},  \,\hat{b}^{\dagger}_{\vec{x}}\}_{-}&=&
 e^{i \vec{p} \cdot \vec{x}} \frac{1}{\sqrt{(2 \pi)^{d-1}}}\,,\qquad,
\{\hat{b}_{\vec{x}},  \,\hat{b}^{\dagger}_{\vec{p}}\}_{-}=
 e^{-i \vec{p} \cdot \vec{x}} \frac{1}{\sqrt{(2 \pi)^{d-1}}}\,,
\end{eqnarray}.
\end{small}

 The internal space of either fermion or boson fields has the finite number of ''basis 
 vectors'', $2^{\frac{d}{2}-1}\times 2^{\frac{d}{2}-1}$, the momentum basis is 
 continuously infinite.\\

The creation operators for either fermions or bosons must be a tensor product, 
$*_{T}$, of both contributions, the ''basis vectors'' describing the internal space of 
fermions or bosons and the basis in ordinary, momentum or coordinate, space. 

The creation operators for a free massless fermion of the energy 
$p^0 =|\vec{p}|$, belonging to a family $f$ and  to a superposition of 
family members $m$  applying on the vacuum state  
$|\psi_{oc}>\,*_{T}\, |0_{\vec{p}}>$ 
can be written as~(\cite{nh2021RPPNP}, Subsect.3.3.2, and the references therein)
 \begin{eqnarray}
\label{wholespacefermions}
{\bf \hat{b}}^{s \dagger}_{f} (\vec{p}) \,&=& \,
\sum_{m} c^{sm}{}_f  (\vec{p}) \,\hat{b}^{\dagger}_{\vec{p}}\,*_{T}\,
\hat{b}^{m \dagger}_{f} \, \,,   
 \end{eqnarray}
where the vacuum state for fermions $|\psi_{oc}>\,*_{T}\, |0_{\vec{p}}> $ 
includes both spaces, the internal part, Eq.(\ref{vaccliffodd}), and the momentum 
part, Eq.~(\ref{creatorp}) (in a tensor product for a starting  single particle state 
with zero momentum, from which one obtains the other single fermion states of the
same ''basis vector'' by the operator  $\hat{b}^{\dagger}_{\vec{p}}$ which pushes 
the momentum by an amount $\vec{p}$~\footnote{
The creation operators and their Hermitian conjugated partners annihilation 
operators in the coordinate representation can be
read in~\cite{nh2021RPPNP} and the references therein:
$\hat{\bf b}^{s \dagger}_{f }(\vec{x},x^0)=
\sum_{m} \,\hat{b}^{ m \dagger}_{f} \,  \int_{- \infty}^{+ \infty} \,
\frac{d^{d-1}p}{(\sqrt{2 \pi})^{d-1}} \, c^{m s }{}_{f}\; 
(\vec{p}) \;  \hat{b}^{\dagger}_{\vec{p}}\;
e^{-i (p^0 x^0- \varepsilon \vec{p}\cdot \vec{x})}
$
~(\cite{nh2021RPPNP}, subsect. 3.3.2., Eqs.~(55,57,64) and the references therein).}). 
\\
The creation operators fulfil the anticommutation relations for the second quantized 
fermion fields
\begin{small}
\begin{eqnarray}
\{  \hat{\bf b}^{s' }_{f `}(\vec{p'})\,,\, 
\hat{\bf b}^{s \dagger}_{f }(\vec{p}) \}_{+} \,|\psi_{oc}> |0_{\vec{p}}>&=&
\delta^{s s'} \delta_{f f'}\,\delta(\vec{p}' - \vec{p})\, |\psi_{oc}> |0_{\vec{p}}>
\,,\nonumber\\
\{  \hat{\bf b}^{s' }_{f `}(\vec{p'})\,,\, 
\hat{\bf b}^{s}_{f }(\vec{p}) \}_{+} \,|\psi_{oc}> |0_{\vec{p}}>&=&0\, . \,
 |\psi_{oc}> |0_{\vec{p}}>
\,,\nonumber\\
\{  \hat{\bf b}^{s' \dagger}_{f '}(\vec{p'})\,,\, 
\hat{\bf b}^{s \dagger}_{f }(\vec{p}) \}_{+}\, |\psi_{oc}> |0_{\vec{p}}>&=&0\, . \,
\,|\psi_{oc}> |0_{\vec{p}}>
\,,\nonumber\\
 \hat{\bf b}^{s \dagger}_{f }(\vec{p}) \,|\psi_{oc}> |0_{\vec{p}}>&=&
|\psi^{s}_{f}(\vec{p})>\,\nonumber\\
 \hat{\bf b}^{s}_{f }(\vec{p}) \, |\psi_{oc}> |0_{\vec{p}}>&=&0\, . \,
 \,|\psi_{oc}> |0_{\vec{p}}>\nonumber\\
 |p^0| &=&|\vec{p}|\,.
\label{Weylpp'comrel}
\end{eqnarray}
\end{small}
The creation operators $  \hat{\bf b}^{s\dagger}_{f }(\vec{p}) )$  and their 
Hermitian conjugated partners annihilation operators  
$\hat{\bf b}^{s}_{f }(\vec{p}) $, creating and annihilating the single fermion 
states, respectively, fulfil when applying on the vacuum state,  
$|\psi_{oc}> *_{T} |0_{\vec{p}}>$, the anticommutation relations for the second quantized 
fermions, postulated by Dirac (Ref.~\cite{nh2021RPPNP}, Subsect.~3.3.1, 
Sect.~5).~\footnote{
 The anticommutation relations of Eq.~(\ref{Weylpp'comrel}) are valid also if we 
 replace  the vacuum state,  $|\psi_{oc}>|0_{\vec{p}}>$, by the Hilbert space of 
Clifford fermions generated by the tensor product multiplication, $*_{T_{H}}$, of 
any number of the Clifford odd fermion states of all possible internal quantum 
numbers and all possible momenta (that is of any number of 
$ \hat {\bf b}^{s\, \dagger}_{f} (\vec{p})$ of any
 $(s,f, \vec{p})$), Ref.~(\cite{nh2021RPPNP}, Sect. 5.).}\\

To write the creation operators for boson fields we must take into account that 
boson gauge fields have the space index $\alpha$, describing the $\alpha$
component of the boson field in the ordinary space~\footnote{
In the  {\it  spin-charge-family} theory  the Higgs's scalars origin 
in the boson gauge fields with the vector index $(7,8)$, Ref.~(\cite{nh2021RPPNP}, Sect.~7.4.1, and the references therein).}.
We  therefore add the space index $\alpha$ as follows
 \begin{eqnarray}
\label{wholespacebosons}
{\bf {}^{I}{\hat{\cal A}}^{m \dagger}_{f \alpha}} (\vec{p}) \,&=& 
\hat{b}^{\dagger}_{\vec{p}}\,*_{T}\, 
{\cal C}^{ m}{}_{f \alpha}\, {}^{I}{\hat{\cal A}}^{m \dagger}_{f} \, \,.                                                                              
 \end{eqnarray}
We treat free massless bosons of momentum $\vec{p}$ and energy $p^0=|\vec{p}|$ 
and of particular ''basis vectors'' ${}^{I}{\hat{\cal A}}^{m \dagger}_{f}$'s which are 
eigenvectors of  all the Cartan subalgebra members~\footnote{
In general the energy eigenstates of bosons  are in superposition of 
${\bf {}^{I}{\hat{\cal A}}^{m \dagger}_{f} }$. One example, which uses the 
superposition of the Cartan subalgebra eigenstates manifesting the $SU(3)\times U(1)$ 
subgroups of the group $SO(6)$,  is presented  in Fig.~\ref{FigSU3U1even}.},
${\cal C}^{ m}{}_{f \alpha}$ carry the space index $\alpha$  of the boson 
field. Creation operators operate on the vacuum state 
$|\psi_{oc_{ev}}>\,*_{T}\, |0_{\vec{p}}> $ with the internal space part
just a constant, $|\psi_{oc_{ev}}>=$ $|\,1>$, and for 
a starting  single boson state with a zero momentum from which one obtains 
the other single boson states with the same ''basis vector'' by the operators 
$\hat{b}^{\dagger}_{\vec{p}}$ which push the momentum by an amount 
$\vec{p}$, making also ${\cal C}^{ m}{}_{f \alpha}$  depending on $\vec{p}$.


For the creation operators for boson fields in  a coordinate
representation we find using  Eqs.~(\ref{creatorp}, \ref{eigenvalue10})
 \begin{eqnarray}
{\bf {}^{I}{\hat{\cal A}}^{m \dagger}_{f \alpha}} 
(\vec{x}, x^0)& =&  \int_{- \infty}^{+ \infty} \,
\frac{d^{d-1}p}{(\sqrt{2 \pi})^{d-1}} \, 
{}^{I}{\hat{\cal A}}^{m \dagger}_{f \alpha}  (\vec{p})\, 
e^{-i (p^0 x^0- \varepsilon \vec{p}\cdot \vec{x})}|_{p^0=|\vec{p}|}\,.
\label{Weylbosonx}
\end{eqnarray}
\vspace{2mm}

To understand what new does the Clifford algebra description of the internal space 
of fermion and boson fields, Eqs.~(\ref{wholespacebosons}, \ref{Weylbosonx}, 
\ref{wholespacefermions}), bring to our understanding of the second quantized 
fermion and boson fields and what new can we learn from this offer, 
we need to relate $\sum_{ab} c^{ab} \omega_{ab \alpha}$ and 
$ \sum_{m f} {}^{I}{\hat{\cal A}}^{m \dagger}_{f} {\cal C}^{m f}_{\alpha}$,
recognizing that ${}^{I}{\hat{\cal A}}^{m \dagger}_{f} {\cal C}^{m f}_{\alpha}$
are eigenstates of the Cartan subalgebra members, while  $\omega_{ab \alpha}$
are not.

The gravity fields, the vielbeins and the two kinds of the spin connection fields,
$f^{a}{}_{\alpha}$, $\omega_{ab \alpha}$, $\tilde{\omega}_{ab \alpha}$, 
respectively, are in the {\it spin-charge-family} theory
(unifying spins, charges and families of fermions and offering not only the 
explanation for all the assumptions of the {\it standard model} but also for the 
increasing number of phenomena observed so far) the only boson fields in 
$d=(13+1)$, observed in $d=(3+1)$ besides as  gravity also as all the other 
boson fields with the Higgs's scalars included~\cite{nd2017}.

We therefore need to relate
\begin{eqnarray}
\label{relationomegaAmf0}
\{\frac{1}{2}  \sum_{ab} S^{ab}\, \omega_{ab \alpha} \} 
\sum_{m } \beta^{m f}\, \hat{\bf b}^{m \dagger}_{f }(\vec{p}) &{\rm relate\, \,to}&
\{ \sum_{m' f '} {}^{I}{\hat{\cal A}}^{m' \dagger}_{f '} \,
{\cal C}^{m' f '}_{\alpha} \}
\sum_{m } \beta^{m f} \, \hat{\bf b}^{m \dagger}_{f }(\vec{p}) \,, \nonumber\\
 &&\forall f \,{\rm and}\,\forall \, \beta^{m f}\,, \nonumber\\
{\bf \cal S}^{cd} \,\sum_{ab} (c^{ab}{}_{mf}\, \omega_{ab \alpha})  &{\rm relate\, \,to}& 
{\bf \cal S}^{cd}\, ({}^{I}{\hat{\cal A}}^{m \dagger}_{f}\, {\cal C}^{m f}_{\alpha})\,, \nonumber\\
&& \forall \,(m,f), \nonumber\\
&&\forall \,\,{\rm Cartan\,\,subalgebra\, \, \, member}  \,{\bf \cal S}^{cd} \,.
\end{eqnarray}
Let be repeated that ${}^{I}{\hat{\cal A}}^{m \dagger}_{f } $ are chosen to be
the eigenvectors of the Cartan subalgebra members, Eq.~(\ref{cartangrasscliff}).
Correspondingly we can relate  a particular ${}^{I}{\hat{\cal A}}^{m \dagger}_{f } 
{\cal C}^{m f }_{\alpha}$ with such a superposition of $\omega_{ab \alpha}$'s
which  is   the eigenvector with  the same values of the Cartan subalgebra members as 
there is a particular ${}^{I}{\hat{\cal A}}^{m \dagger}_{f } {\cal C}^{m f }_{\alpha}$. 
We can do this in two ways:\\
 {\bf i.} $\;\;$ Using the first relation in Eq.~(\ref{relationomegaAmf0}).  On the left 
hand side of this relation ${S}^{ab}$'s apply  on $ \hat{b}^{m \dagger}_{f} $  part of 
 $ \hat{\bf b}^{m \dagger}_{f }(\vec{p}) $.
On the right hand side ${}^{I}{\hat{\cal A}}^{m \dagger}_{f }$ apply as well on  the
same ''basis vector''  $ \hat{b}^{m \dagger}_{f} $. \\
  {\bf ii.} $\;\;$ Using  the second relation, in which  ${\bf \cal S}^{cd}$ apply  on 
   the left hand side on  $\omega_{ab \alpha}$'s
\begin{eqnarray}
\label{sonomega}
 \, {\bf \cal S}^{cd} \,\sum_{ab}\, c^{ab}{}_{mf}\, \omega_{ab \alpha}
 &=& \sum_{ab}\, c^{ab}{}_{mf}\, i \,(\omega_{cb \alpha} \eta^{ad}- 
\omega_{db \alpha} \eta^{ac}+ \omega_{ac \alpha} \eta^{bd}-
\omega_{ad \alpha} \eta^{bc}),
\end{eqnarray}
on  each $ \omega_{ab \alpha}$ separately; $c^{ab}{}_{mf}$ are constants to be 
determined from the second relation, where  on the right hand side of this relation
${\bf \cal S}^{cd} (= S^{cd}+ \tilde{S}^{cd})$ apply on the ''basis vector'' 
${}^{I}{\hat{\cal A}}^{m \dagger}_{f }$ of the corresponding gauge field.

Let us conclude this section by pointing out that either the Clifford odd ''basis vectors''
$\hat{b}^{m \dagger}_{f}$ or the Clifford even ''basis vectors'' 
${}^{i}{\hat{\cal A}}^{m \dagger}_{f}, i=(I,II) $ have in any even $d$ 
$2^{\frac{d}{2}-1}$ $\times 2^{\frac{d}{2}-1}$ members, while $\omega_{ab \alpha}$
as well as $\tilde{\omega}_{ab \alpha}$ have each for each $\alpha$ $\frac{d}{2}(d-1)$
members. It is needed to find out what new does this difference bring into 
the - unifying theories of the Kaluza-Klein theories are.

\section{Short overview   and achievements of {\it spin-charge-family} theory}
\label{SCFT}

\vspace{2mm}

The {\it spin-chare-family} theory~\cite{norma92,norma93,IARD2016,%
n2014matterantimatter,nd2017,n2012scalars,JMP2013,normaJMP2015,nh2017,%
nh2018} is a kind of the Kaluza-Klein theories~\cite{KaluzaKlein,%
Witten,Duff,App,SapTin,Wetterich,zelenaknjiga,mil,nd2017} since it is built on
the assumption that the dimension of  space-time is $\ge (13 +1)$~\footnote{
$d=(13+1)$  is the smallest dimension for which the subgroups of the group 
$SO(13,1)$ offer the description of spins and charges of fermions assumed 
by the {\it standard model} and correspondingly also of boson gauge fields.},
and that the only interaction among fermions is the gravitational one (vielbeins, 
the gauge fields of momenta, and two kinds of the spin connection fields, the 
gauge fields of $S^{ab}$ and of $\tilde{S}^{ab}\;$%
\footnote{If there are no fermions present both spin connection fields are 
expressible with vielbeins~(\cite{nh2021RPPNP}, Eq.~(103)).}).

This theory assumes as well that the internal space of fermion and boson fields 
are described by the Clifford odd and Clifford even algebra, respectively~\footnote{
Fermions and bosons internal spaces are assumed to be superposition of odd 
products of $\gamma^{a}$'s (fermion fields) or of even products of 
$\gamma^{a}$'s (boson fields) what offers the explanation for 
the second quantized postulates of Dirac~\cite{Dirac}. The ''basis 
vectors'' of the internal spaces  namely determine anticommutativity or 
commutativity of the corresponding creation and annihilation operators.}. 

The theory is offering the explanation for all the assumptions of the {\it standard 
model}, unifying not only charges, but also spins, charges  and families, and 
consequently offering the explanation for the appearance of families of quarks and 
leptons and antiquarks and antileptons, of vector gauge fields, of Higgs's scalar 
field and the Yukawa couplings, for the differences in masses among quarks and 
leptons for the matter-antimatter asymmetry in the universe, for the dark 
matter, making several predictions. 

The {\it spin-charge-family} theory shares with the Kaluza-Klein like theories
 their weak points, like: {\;\;\;\bf a.} Not yet solved the quantization 
problem of the gravitational field~\footnote{The description of the internal 
space of fermions and bosons as superposition of odd (for fermion fields) or
even (for boson fields) products of the Clifford objects $\gamma^{a}$'s  
seems very promising in looking for a new way to second quantization of all
fields, with gravity included, as discussed in this talk.}. $\;\;\;$
{\;\;\;\bf b.}~The spontaneous brake of the starting symmetry which would 
at low energies manifest the observed almost massless fermions~\cite{Witten}.   
The spontaneously break of the starting symmetry of $SO(13+1)$ with the 
condensate of the two right handed neutrinos  (with the family quantum 
numbers of the group of four families, which does not include the observed 
three families~(\cite{prd2018},~Table~III),~(\cite{nh2021RPPNP},~Table~6)
bringing masses of the scale $\propto 10^{16}$ GeV or higher to all the 
vector and scalar gauge fields,  which interact with the 
condensate~\cite{n2014matterantimatter} is promissing to show the right 
way ~\cite{NHD}. 

The scalar fields (scalar fields are the spin connection fields with the space index 
$\alpha$  higher than $(0,1,2,3)$)  with the space index $(7,8)$ offer, after 
gaining constant non zero vacuum values, the explanation for the Higgs's 
scalar and the Yukawa couplings. 
They namely determine the mass matrices of quarks and leptons and 
antiquarks and antileptons. In Refs.~\cite{nd2017,IARD2020} it is pointed 
out that the spin connection gauge fields do manifest in $d=(3+1)$ as the 
ordinary gravity and all the observed vector and scalar gauge fields.

The {\it spin-charge-family} theory  assumes a simple starting action for 
second quantized massless fermion and the corresponding gauge 
boson fields in $d=(13+1)$-dimensional space,  presented in 
Eq.~(\ref{wholeaction}).

The  fermion part of the action, Eq.~(\ref {wholeaction}), can be rewritten in 
the way that it manifests in $d=(3+1)$ in the low energy regime before the 
electroweak break by the {\it standard model} postulated properties of: 
$\;\;$ {\bf i.} Quarks and leptons and antiquarks and antileptons with the spins, 
handedness, charges and family quantum numbers. Their internal space is 
described by the Clifford odd ''basis vectors'' which are eigenvectors of the 
Cartan subalgebra of $S^{ab}$ and $\tilde{S}^{ab}$, Eqs.~(\ref{cartangrasscliff},
\ref{so64 5+1}, \ref{so1+3 5+1}). 
$\;\;$ {\bf ii.} Couplings of fermions to the vector gauge fields, which are the 
superposition of gauge fields $\omega^{st}{}_{\alpha} $, Sect.~6.2 in Ref.~\cite{nh2021RPPNP}, with the space index $\alpha=(0,1,2,3)$ and with the
charges determined by the Cartan subalgebra of $S^{ab}$ and $\tilde{S}^{ab}$ 
manifesting the symmetry of space $(d-4)$, and  to the  scalar gauge 
fields~\cite{IARD2016,normaBled2020,JMP2013,normaJMP2015,pikanorma,%
pikanorma2005,norma92,norma93,gmdn2008,gn2009,gn2013,IARD2020} 
with the space index $\alpha \ge5$
and the charges determined  by the Cartan subalgebra of $S^{ab}$ 
and $\tilde{S}^{ab}$ (as explained in the case of the vector gauge fields), and 
which are superposition of either $\omega^{st}{}_{\alpha} $ or 
$\tilde{\omega}^{ab}{}_{\alpha} $, 
\begin{eqnarray}
\label{faction}
{\mathcal L}_f &=&  \bar{\psi}\gamma^{m} (p_{m}- \sum_{A,i}\; g^{Ai}\tau^{Ai} 
A^{Ai}_{m}) \psi + \nonumber\\
               & &  \{ \sum_{s=7,8}\;  \bar{\psi} \gamma^{s} p_{0s} \; \psi \} +
 \nonumber\\ 
& & \{ \sum_{t=5,6,9,\dots, 14}\;  \bar{\psi} \gamma^{t} p_{0t} \; \psi \}
\,, 
\end{eqnarray}
where $p_{0s} =  p_{s}  - \frac{1}{2}  S^{s' s"} \omega_{s' s" s} - 
                    \frac{1}{2}  \tilde{S}^{ab}   \tilde{\omega}_{ab s}$, 
$p_{0t}   =    p_{t}  - \frac{1}{2}  S^{t' t"} \omega_{t' t" t} - 
                    \frac{1}{2}  \tilde{S}^{ab}   \tilde{\omega}_{ab t}$,                    
with $p_{0s}= e_{s}^{\alpha} p_{0\alpha}$,  $m \in (0,1,2,3)$, 
$s \in (7,8),\, (s',s") \in (5,6,7,8)$, $(a,b)$ (appearing in
 $\tilde{S}^{ab}$) run within  either $ (0,1,2,3)$ or $ (5,6,7,8)$, $t$ runs 
$ \in (5,\dots,14)$, 
$(t',t")$ run either $ \in  (5,6,7,8)$ or $\in (9,10,\dots,14)$. 
The spinor function $\psi$ represents all family members of all the 
$2^{\frac{7+1}{2}-1}=8$ families.

\vspace{3mm}

$\;\;$ The first line of Eq.~(\ref{faction}) determines in $d=(3+1)$ the kinematics 
and dynamics of fermion fields coupled to the vector gauge fields~\cite{nd2017,normaJMP2015,IARD2016}. 
The vector gauge fields are the superposition of the spin connection fields 
$\omega_{stm}$, $m=(0,1,2,3)$, $(s,t)=(5,6,\cdots,13,14)$, and are the
gauge fields of $S^{st}$, Subsect.~(6.2.1) of Ref.~\cite{nh2021RPPNP}.

The reader can find in Sect.~6 of Ref.~\cite{nh2021RPPNP} a quite detailed overview
of the properties which the massless fermion and boson fields appearing in the 
simple starting action, Eq.~(\ref{wholeaction}), (the later only as gravitational
fields) manifest in $d=(3+1)$ as all the observed fermions --- quarks and 
leptons and antiquarks and antileptons in each family --- appearing in twice four 
families, with the 
lower four families including the observed three families of quarks and leptons and 
antiquarks and antileptons. The higher four families offer the explanation for the dark 
matter.

Table~5 and Eq.~(110) of Ref.~\cite{nh2021RPPNP} explain that the scalar fields
with the space index $\alpha =(7,8)$ carry the weak charge $\tau^{13}=
\pm \frac{1}{2}$ and the hyper charge $Y=\mp \frac{1}{2}$, just as assumed by 
the  {\it standard model}.

Masses of families of quarks and leptons are determined by the superposition of 
the scalar fields, Eq.~(108-120) of Ref.~\cite{nh2021RPPNP}, appearing in two 
groups, each of them manifesting
the symmetry $SU(2)$$\times SU(2)$ $ \times U(1)$~\footnote{
The assumption that the symmetry  $SO(13,1)$ first breaks into $SU(3)\times $
$U(1) \times SO(7,1)$ makes that quarks and leptons distinguish only in the 
part  $SU(3)\times U(1) $, while the $SO(7,1)$ part is identical  separately for quarks 
and leptons and separately for antiquarks and antileptons. Table 7 of 
Ref.~\cite{nh2021RPPNP}, 
presenting one family, which includes quarks and leptons and antiquarks and 
antileptons, manifests these properties. The $\omega_{ab \alpha}$, with the 
space index $(7,8)$ carry with respect to the flat index $ab$ only quantum 
numbers $Q,Y,\tau^4$, ($Q$ $( =  \tau^{13} + Y)$,$\tau^{13}$ 
$(= \frac{1}{2}({\cal S}^{56} - {\cal S}^{78})$, $Y$ 
$(= \tau^{4} + \tau^{23})$ and $\tau^4=-\frac{1}{3} ({\cal S}^{9\,10}
 +  {\cal S}^{11\,12} +{\cal S}^{13\,14})$, the flat index $(ab)$ of 
 $\tilde{\omega}_{ab \alpha}$,  with the space index $(7,8)$, includes
 all ($0,1,\dots,8$) correspondingly forming the symmetry 
 $SU(2)$$\times SU(2)$ $ \times U(1)$.}.

 The scalar gauge fields with the space index $(7,8)$ determine correspondingly 
 the symmetry of mass matrices of quarks and leptons~(\cite{nh2021RPPNP},  
 Eq.~(111)) which appear in two groups as the scalar fields do.  In Table 5 in 
 Ref.~\cite{nh2021RPPNP}) the symmetry $SU(2)\times SU(2)\times U(1 )$ for 
 each of the two groups is presented and explained. 
 
 Although  spontaneous symmetry braking of the starting symmetry has not (yet
 consistently enough)  been studied and the coupling constants of the scalar 
 fields among themselves  and with quarks and leptons are not yet known, the 
 known symmetry of mass  matrices, presented in Eq.~(111) of 
 Ref.~\cite{nh2021RPPNP}, enables to 
 determine parameters of mass matrices from the measured data of the $3\times3$
 sub mixing matrices and the masses of the measured three families of quarks 
 and leptons. 
 
 Although the known $3\times 3$ submatrix of the unitary $4\times 4$ matrix
 enables to determine $4\times 4$ matrix, the measured $3\times 3$ mixing 
 sub matrix is even for quarks far accurately enough measured, so that we only 
 can  predict the matrix elements of the $4\times4$ mixing matrix  for quarks 
if assuming that masses (times $c^2$) of the fourth family quarks are heavy 
enough, that is above one TeV~\cite{mdn2006,gn2009}.  
The new measurements of the matrix 
elements among the observed $3$ families agree  better with the predictions
obtained by the {\it sspin-charge-family} theory than the old measurements.
The reader can find predictions in Refs.~(\cite{gn2013,gn2014}) and the 
overview in Ref.~(\cite{nh2021RPPNP}, Subsect.~7.3.1).

The upper group of four families offers the explanation for the Dark matter,
to which the quarks and leptons from the (almost) stable of the upper four families 
mostly contribute. The reader can find the report on this proposal for the Dark matter 
origin in Ref.~\cite{gn2009} and a short overview in Subsect.~7.3.1 of 
\cite{nh2021RPPNP}, where the appearance, development and properties of the 
Dark matter are  discussed. The upper four families predict nucleons of very 
heavy quarks with the nuclear force among nucleons which is correspondingly 
very different from the known one~\cite{gn2009,nm2015}.

\vspace{2mm}

Besides the scalar fields with the space index $\alpha=(7,8)$, which manifest
in $d=(3+1)$ as scalar gauge fields with the weak and hyper charge $\pm \frac{1}{2}$ 
and $\mp \frac{1}{2}$, respectively, and which gaining at low energies constant 
values  make families of quarks and leptons and the weak gauge field massive, there 
are in the starting action, Eqs.~(\ref{wholeaction}), additional scalar gauge
fields with the space index $\alpha=(9,10,11,12,13,14)$. They are with respect
to the space index $\alpha$ either triplets or antitriplets causing transitions from 
antileptons into quarks and from antiquarks into quarks and back. 

Their properties are presented in Ref.~\cite{n2014matterantimatter} and briefly in 
Table 9 and Fig. 1 of Ref.~\cite{nh2021RPPNP}.

Concerning this second point we proved on the toy model of $d=(5+1)$ that the break of 
symmetry can lead to (almost) massless fermions~\cite{NHD}. 

In $d=(3+1)$-dimensional space --- at low energies --- the gauge gravitational fields 
manifest as the observed vector gauge fields~\cite{nd2017},  which can be quantized 
in the usual  way. 
The author is in mean time trying to find out (together with the collaborators)  
how far can the {\it spin-charge-family} theory --- starting in $d=(13+1)$-dimensional space with
a simple and "elegant" action, Eq.~(\ref{wholeaction}) ---  reproduce in $d=(3+1)$ the observed 
properties of  quarks and leptons~\cite{IARD2016,n2014matterantimatter,nd2017,n2012scalars,%
JMP2013,normaJMP2015,nh2017,nh2018}, the observed vector gauge fields, the  scalar field 
and thge Yukawa couplings, the appearance 
of the dark matter and of the matter-antimatter asymmetry, as well as the other
open questions, connecting elementary fermion and boson fields and cosmology. 

The work done so far on the {\it spin-charge-family} theory seems promising.

\section{Conclusions}
\label{conclusions}

\vspace{2mm}

In the {\it spin-charge-family} theory~\cite{norma92,norma93,IARD2016,%
n2014matterantimatter,nd2017,n2012scalars,JMP2013,normaJMP2015,nh2017,%
nh2018,nh2021RPPNP} 
the Clifford odd algebra is used to describe the internal space of fermion fields. The 
Clifford odd ''basis vectors'' --- the  superposition of odd products of $\gamma^a$'s 
--- in a tensor 
product with the basis in ordinary space form the creation and annihilation operators, 
in which the anticommutativity of the ''basis vectors'' is transferred  to the creation 
and annihilation operators for fermions, offering the explanation for the second 
quantization postulates for fermion fields. 

The  Clifford odd ''basis vectors'' have all the properties of fermions: Half integer 
spins with respect to the Cartan subalgebra members of the Lorentz algebra in 
the internal space of fermions in even dimensional spaces ($d=2(2n+1)$ or 
$d=4n$), as discussed in  Subsects.~(\ref{basisvectors},
\ref{secondquantizedfermionsbosons}).\\
With respect to the subgroups of the $SO(d-1, 1)$ group the Clifford odd ''basis 
vectors'' appear in the fundamental representations, as illustrated in 
Subsects.~\ref{cliffordoddevenbasis5+1}. 

In this article it is demonstrated that the Clifford even algebra is offering the description 
of the internal space of boson fields. The Clifford even ''basis vectors'' --- the  
superposition of even products of $\gamma^a$'s --- in a tensor product 
with the basis in  ordinary space form the creation and annihilation operators which
manifest the commuting properties of the second quantized boson fields, offering 
explanation for the second quantization postulates for boson fields~\cite{n2021SQ}.
The Clifford even ''basis vectors'' have all the properties of bosons: Integer spins with 
respect to the Cartan subalgebra members of the Lorentz algebra in the internal space 
of bosons, as discussed in  Subsects.~\ref{basisvectors}.\\

With respect to the subgroups of the $SO(d-1, 1)$ group the Clifford even ''basis 
vectors'' manifest the  adjoint representations, as illustrated in
Subsect.~\ref{cliffordoddevenbasis5+1}.

There are two kinds of anticommuting algebras~\cite{norma93}: The Grassmann
algebra, offering in $d$-dimensional space $2\,.\, 2^d$ operators ($2^d$ $\,\theta^a$'s 
and $2^d$ $\frac{\partial}{\partial \theta_a}$'s, Hermitian conjugated to each other, Eq.~(\ref{thetaderher0})),  and the two Clifford subalgebras, each with $2^d$ operators 
named $\gamma^a$'s and $\tilde{\gamma}^a$'s, respectively, \cite{norma93,nh02,nh03}, Eqs.~(\ref{thetaderanti0}-\ref{gammatildeantiher}).

The operators in each of the two  Clifford subalgebras  appear in two groups of 
$2^{\frac{d}{2}-1}\times $ $2^{\frac{d}{2}-1}$ of  the Clifford odd  operators 
(the odd products of either $\gamma^a$'s in one subalgebra or of 
$\tilde{\gamma}^a$'s in the other subalgebra),  which are Hermitian conjugated 
to each other: In each Clifford odd group of any of the two subalgebras there appear 
$2^{\frac{d}{2}-1}$ irreducible representation each with the $2^{\frac{d}{2}-1}$
members and the group of their Hermitian conjugated partners.

There are as well the Clifford even operators (the even products of either 
$\gamma^a$'s in one subalgebra or of $\tilde{\gamma}^a$'s in another 
subalgebra) which  again appear in two groups of $2^{\frac{d}{2}-1}\times $ 
$2^{\frac{d}{2}-1}$ members each. In the case of the Clifford even objects the 
members of each group of $2^{\frac{d}{2}-1}\times $ $2^{\frac{d}{2}-1}$ 
members have the Hermitian conjugated partners within the same group, ~Subsect.~\ref{basisvectors}, Table~\ref{Table Clifffourplet.}.

The Grassmann algebra operators are  expressible with the operators of the two Clifford subalgebras and opposite,~Eq.~(\ref{clifftheta1}). The two Clifford subalgebras are 
independent of each other, Eq.~(\ref{gammatildeantiher}), forming two independent 
spaces.

Either the Grassmann algebra~\cite{ND2018Grass,n2019PIPII} or the two Clifford 
subalgebras can be used to describe the internal space of anticommuting objects, 
if the superposition of odd products of operators 
($\theta^a$'s or $\gamma^a$'s, or $ \tilde{\gamma}^a$'s) are used to describe the 
internal space of these objects. The commuting objects must be superposition of 
even products of operators  ($\theta^a$'s or $ \gamma^a$'s or $\tilde{\gamma}^a$'s).

\vspace{2mm}

No integer spin anticommuting objects have been observed so far, and to describe the 
internal space of the so far observed fermions only one of the two Clifford odd 
subalgebras are needed. 

\vspace{2mm}

The problem can be solved by reducing  the two Clifford sub algebras to only one, the one 
(chosen to be) determined by $\gamma^{ab}$'s. The decision  
that $ \tilde{\gamma}^a$'s  apply  on $ \gamma^a$ as follows: 
$\{ \tilde{\gamma}^a B =(-)^B\, i \, B \gamma^a\}\, |\psi_{oc}>$, 
Eq.~(\ref{tildegammareduced}), 
(with $(-)^B = -1$, if $B$ is a function of an odd products of $\gamma^a$'s,
otherwise $(-)^B = 1$) enables that  $2^{\frac{d}{2}-1}$ irreducible representations  
of $S^{ab}= \frac{i}{2}\, \{\gamma^a\,,\, \gamma^b\}_{-}$ (each with the  
$2^{\frac{d}{2}-1}$ members) obtain the family quantum numbers determined by  
$\tilde{S}^{ab}= \frac{i}{2}\, \{\tilde{\gamma}^a\,,\,\tilde{\gamma}^b\}_{-}$.
 
\vspace{2mm}

The decision to use in the {\it spin-charge-family} theory in $d=2(2n +1)$, $n\ge 3$
($d\ge (13+1)$ indeed),
the superposition of the odd products of the Clifford algebra elements $\gamma^{a}$'s 
to describe  the internal space of fermions  which interact with the gravity only 
(with the vielbeins, the gauge fields of momenta, and the two kinds of the spin 
connection fields, the gauge fields of  $S^{ab}$ and $\tilde{S}^{ab}$, respectively), Eq.~(\ref{wholeaction}), offers not 
only the explanation for all the assumed properties of fermions and bosons in 
the {\it standard model}, with the appearance of  the families of quarks and leptons 
and antiquarks and antileptons~(\cite{nh2021RPPNP} and the references therein) and 
of the corresponding vector gauge fields  and the Higgs's scalars included~\cite{nd2017},  
but also for the appearance of the dark matter~\cite{gn2009} in the universe, for the explanation of the matter/antimatter asymmetry in the 
universe~\cite{n2014matterantimatter},  and for several other observed phenomena, 
making several predictions~\cite{pikanorma2005,gmdn2007,gmdn2008,gn2013}. 

\vspace{2mm}

Recognition that the use of the superposition of the even products of the Clifford algebra elements $\gamma^{a}$'s to describe the internal space  of boson fields, what appear
to manifest all the 
properties of the observed boson fields, as demonstrated in this articles, makes clear
that the Clifford algebra offers not only the explanation for the postulates of the second quantized anticommuting fermion fields but also for the postulates of the second 
quantized boson fields.

The relations in Eq.~(\ref{relationomegaAmf0}) 
\begin{eqnarray}
\label{relationomegaAmf01}
\{\frac{1}{2}  \sum_{ab} S^{ab}\, \omega_{ab \alpha} \} 
\sum_{m } \beta^{m f}\, \hat{\bf b}^{m \dagger}_{f }(\vec{p}) &{\rm relate\,\, to}&
\{ \sum_{m' f '} {}^{I}{\hat{\cal A}}^{m' \dagger}_{f '} \,
{\cal C}^{m' f '}_{\alpha} \}
\sum_{m } \beta^{m f} \, \hat{\bf b}^{m \dagger}_{f }(\vec{p}) \,, \nonumber\\
 &&\forall f \,{\rm and}\,\forall \, \beta^{m f}\,, \nonumber\\
{\bf \cal S}^{cd} \,\sum_{ab} (c^{ab}{}_{mf}\, \omega_{ab \alpha})  &{\rm relate\,\, to}& 
{\bf \cal S}^{cd}\, ({}^{I}{\hat{\cal A}}^{m \dagger}_{f}\, {\cal C}^{m f}_{\alpha})\,, \nonumber\\
&& \forall \,(m,f), \nonumber\\
&&\forall \,\,{\rm Cartan\,\,subalgebra\, \, \, member}  \,{\bf \cal S}^{cd} \,,
\nonumber
\end{eqnarray}
offers the possibility to replace the covariant derivative 
$ p_{0 \alpha }$
  $$p_{0\alpha} = p_{\alpha}  - \frac{1}{2}  S^{ab} \omega_{ab \alpha} - 
                    \frac{1}{2}  \tilde{S}^{ab}   \tilde{\omega}_{ab \alpha} 
                    \quad \quad \quad\quad\;$$
in Eq.~(\ref{wholeaction}) with 

$$ p_{0\alpha}  = p_{\alpha}  - 
\sum_{m f}   {}^{I}{ \hat {\cal A}}^{m \dagger}_{f}
{}^{I}{\cal C}^{m}_{f \alpha}   - 
 \sum_{m f} {}^{I}{\hat{\widetilde{\cal A}}}^{m \dagger}_{f}\,
{}^{I}{\widetilde{\cal C}}^{m}_{f \alpha}\,, $$ 
\noindent
where the relation among ${}^{I}{\hat{\widetilde{\cal A}}}^{m \dagger}_{f}
{}^{I}{\widetilde{\cal C}}^{m}_{f \alpha}$ and 
${}^{II}{\hat{\widetilde{\cal A}}}^{m \dagger}_{f}\,
{}^{II}{\widetilde{\cal C}}^{m}_{f \alpha}$ with respect to $\omega_{ab \alpha}$
 and $\tilde{\omega}_{ab \alpha}$, not discussed directly in this article, needs additional
 study and explanation. 

Although the properties of the Clifford odd and even ''basis vectors'' and correspondingly 
of the creation and annihilation operators for fermion and boson fields are, hopefully,  
clearly demonstrated in this article, yet the proposed way of the second quantization 
of fields, the fermion and the boson ones, needs further study to find out what new can 
the description of the internal space of fermions and bosons bring in understanding of 
the second quantized fields. 

It looks like that this study, showing up  that the Clifford algebra can be used to 
describe  the internal spaces of fermion and boson fields in an equivalent way, offering 
correspondingly the explanation for the second quantization postulates for fermion and 
boson fields, is opening the new insight into the quantum field theory,
since studies of the interaction of fermion fields with boson fields and of boson fields with
boson fields so far looks very promising.

 The study of properties  of the second quantized boson fields, the internal space of 
  which is described by the Clifford even algebra, has just started and needs further
consideration.

\subsection{Acknowledgment} 
The author thanks Department of Physics, FMF, University of Ljubljana, Society of Mathematicians, Physicists and Astronomers of Slovenia,  for supporting the research on the {\it spin-charge-family} theory by offering the room and computer facilities and Matja\v z Breskvar of Beyond Semiconductor for donations, in particular for the annual workshops entitled "What comes beyond the standard models". 


   \end{document}